\documentclass[preprint]{elsarticle}
\usepackage[utf8]{inputenc}
\makeatletter
\def\ps@pprintTitle{%
 \let\@oddhead\@empty
 \let\@evenhead\@empty
 \def\@oddfoot{\reset@font\hfil\thepage\hfil}
 \let\@evenfoot\@oddfoot
}
\usepackage{epstopdf}

\usepackage{graphics}
\usepackage{epsfig}
\usepackage{amsmath}
\usepackage{amsfonts}
\usepackage{cases}
\usepackage{amsthm}
\usepackage{color}
\usepackage{makecell}
\usepackage{multirow}
\usepackage{caption}
\usepackage{subcaption}
\usepackage{natbib}
\usepackage{placeins}
\usepackage{epstopdf}
\usepackage{subcaption}

\usepackage{pifont}
\newcommand{\cmark}{\ding{51}} 
\newcommand{\xmark}{\ding{55}} 

\usepackage{booktabs}
\usepackage{siunitx}
\usepackage{comment}
\catcode`@=11
\@addtoreset{equation}{section}
\renewcommand\theequation{\thesection.\@arabic\c@equation}
\catcode`@=12

\usepackage{float}

\usepackage{booktabs,makecell,adjustbox,graphicx,pifont,array}
\renewcommand{\arraystretch}{1.15}
\newcolumntype{C}[1]{>{\centering\arraybackslash}p{#1}} 

\usepackage{geometry}
\usepackage{color}

\newcommand{\blue}[1]{{\color{blue} #1}}

\numberwithin{equation}{section}

\usepackage{color}

\usepackage{graphicx} 

\usepackage{booktabs}   
\usepackage{pifont}     

\usepackage{color}

\usepackage{graphicx} 
\usepackage{indentfirst}

\usepackage[plainpages=false]{hyperref}
\hypersetup{colorlinks=true,citecolor=blue,filecolor=black,linkcolor=red,urlcolor=green}

\begin{document}
	
	\begin{frontmatter}

         \title{ \bf CBINNS: Cancer Biology-Informed Neural Network for Unknown Parameter Estimation and Missing Physics Identification
         \tnoteref{mytitlenote}}

          \author[nitt]{Bishal Chhetri }
    
        \author[nitt]{B.V. Rathish Kumar}


%
		\address[nitt]{Department of Mathematics and Statistics, \\
  Indian Institute of Technology Kanpur, Kanpur - 208016, Uttar Pradesh, India}

 \begin{abstract}

The dynamics of tumor-immune interactions within a complex tumor microenvironment are typically modeled using a system of ordinary differential equations or partial differential equations.  These models introduce some unknown parameters that need to be estimated accurately and efficiently from the limited and noisy experimental data.  Moreover, due to the intricate biological complexity and limitations in experimental measurements, tumor-immune dynamics are not fully understood, and therefore, only partial knowledge of the underlying physics may be available, resulting in unknown or missing terms within the system of equations.  Thus, there are twofold challenges in modeling tumor dynamics: (i) accurate estimation of model parameters and (ii) discovery of the mathematical equations governing the physical and biological systems. These types of problems are referred to as gray-box identification areas, where both experimental data and partial system knowledge are used to recover unknown parameters and missing components. The use of neural networks to solve and analyze complex physical, biological, and engineering systems has gained significant attention in recent years. In this study, we develop a cancer biology-informed neural network model (CBINN) to infer the unknown parameters in the system of equations as well as to discover the missing physics from sparse and noisy measurements. We test the performance of the CBINN model on three distinct nonlinear compartmental tumour–immune models and evaluate its robustness across multiple synthetic noise levels.  By harnessing these highly nonlinear dynamics, our CBINN framework effectively estimates the unknown model parameters and uncovers the underlying physical laws or mathematical structures that govern these biological systems, even from scattered and noisy measurements. The models chosen here represent the dynamic patterns commonly observed in compartmental models of tumor-immune interactions, thereby validating the generalizability and efficacy of our methodology. This work provides valuable guidance for researchers addressing inverse problems and gray-box identification challenges in complex dynamical systems.


 \end{abstract}
        
\begin{keyword}
Tumor Microenvironment; Parameter Estimation; Missing Physics Identification; Physics-Informed Neural Networks (PINNs); Cancer Biology Informed Neural Networks (CBINN); Structural Identifiability; Fisher Information Matrix; Practical Identifiability
\end{keyword}

\end{frontmatter}
 
\section{Introduction and Motivation}

Cancer is a disease of abnormal cell growth. The uncontrolled growth results in the rapid accumulation of abnormal cells that invade adjoining parts of the body and spread, or metastasis, to other organs affecting the organs and their functions leading to death ultimately. Most of the body’s cells have specific functions and fixed lifespans, undergoing a natural process called apoptosis, where they are programmed to die when no longer needed. This allows the body to replace old or damaged cells with new, functional ones. Cancerous cells, however, bypass these regulatory mechanisms. They lack the signals to stop dividing and to undergo apoptosis, leading to uncontrolled growth and tumor formation \cite{loeb1938causes}. Localized tumors can often be effectively treated with surgery or radiation, resulting in high survival rates. However, as cancer progresses and invades surrounding tissues or enters the bloodstream, it becomes much more challenging to treat, significantly reducing the patient's chances of survival. \cite{sunny2008cancer}. \\

A tumor is not just a conglomerate of malignant cells, but is a heterogeneous collection of many cell types, secreted factors, and extracellular matrix \cite{anderson2020tumor}. Immune cells,  fibroblasts, lymphocytes, blood vessels, bone marrow-derived inflammatory cells, and signaling molecules constitute tumor environment and the interactions between these cells and molecules determines the development and progression of cancer \cite{arneth2019tumor}.  Each of these cells have specific roles and functions, contributing to the tumor microenvironment (TME) by either suppressing or promoting cancer growth. The tumor-suppressing or growth-promoting roles of cells in the TME are mediated through complex signaling networks that include cytokines, chemokines, growth factors, inflammatory enzymes, and matrix remodeling proteinases \cite{balkwill2012tumor}.  
The dynamic and complex interactions between different components of the tumor microenvironment are usually modeled by a system of nonlinear ordinary differential equations (ODEs), which describes the time evolution of the concentrations of interacting components. These models involve several unknown parameters that need to be estimated accurately and efficiently to ensure biological relevance and predictive power, for example, the tumor proliferation rate, the immune cell recruitment rate, and the immune-mediated tumor killing rate. Thus, one central challenge in computational modeling of these systems is the estimation of model parameters  and the prediction of model dynamics. Hence, a lot of attention has been given to the problem of parameter estimation in the systems biology community. In particular, extensive research has been conducted on the applications of different optimization techniques, such as linear and nonlinear least-squares fitting \cite{mendes1998non}. genetic algorithms \cite{moles2003parameter}, and Bayesian method \cite{wilkinson2007bayesian}. Some additional estimation methods have been discussed in \cite{engelhardt2016learning, engelhardt2017bayesian}, but in these examples the number of observable variables required is almost one half of the number of total model variables.\\

Another challenge in systems biology is discovering of mathematical equations that govern physical and biological systems from sparse observations. The tumor microenvironment (TME) is a complex continuously evolving dynamic system that plays a crucial role in the progression and treatment of cancer.  The intricate biological interactions, mediated through networks of cytokines, chemokines, and growth factors, create a complex dynamic and evolving system that is not fully understood.  Given this biological complexity and the limitations of experimental techniques, our understanding of tumor-immune dynamics remains incomplete.  The complexity and heterogeneity inherent in tumor biology give rise to unknown or partially characterized processes, which manifest as missing or incomplete components in mathematical models. \\

 The use of neural networks to solve and analyze complex physical, biological, and engineering systems has gained significant attention in recent years. Using the ability of deep neural networks (DNN) to act as universal function approximators \cite{hornik1989multilayer}, combined with the power of automatic differentiation \cite{baydin2018automatic}, researchers have addressed a wide range of challenging problems in computational science. By embedding physical constraints such as symmetries, invariances, and conservation laws directly into the architecture or loss function, these networks are designed to respect the underlying governing laws, typically represented by time-dependent and nonlinear partial differential equations (PDEs). This simple yet powerful framework forms the foundation of Physics-Informed Neural Networks (PINNs) and has enabled the development of data-efficient learning algorithms, novel numerical solvers for PDEs, and robust data-driven tools for system identification and model inversion. PINNs integrate data-driven learning with physical laws embedded as differential equations, enabling simultaneous model fitting and discovery of unknown dynamics. This approach is especially valuable when dealing with sparse, noisy, or partial data, as it enforces consistency with known physics while allowing flexibility to identify missing terms. In \cite{raissi2019physics}, two types of problem with data-driven solutions and data-driven discovery of partial differential equations are discussed using physics-informed neural networks for both continuous- and discrete-time models. This research has drawn significant attention to the PINN approach for solving both forward and inverse problems. The authors in \cite{yazdani2020systems} developed a systems-biology-informed deep learning algorithm to infer the parameters and predict the hidden dynamics of cell apoptosis, ultradian endocrine model and yeast glycolysis model.  Physics-informed neural networks frameworks have been shown to be very effective in discovering missing physics. In  \cite{ahmadi2024ai}, the authors present a new physics informed
framework for missing physics identification (gray-box)
in the field of systems biology. Several other studies have been carried out to retrieve governing equations from data.  In \cite{brunton2016discovering}, a new idea based on sparse regression and compressed sensing is proposed for dynamical system discovery problem. The symbolic regression based method is proposed by authors in \cite{udrescu2020ai} to represents an unknown function based on a given dataset. Random Projection Neural Networks (RPNNs) in combination with symbolic regression has been studied and are found to have great efficiency in solving forward problems of stiff ODEs, outperforming traditional solvers \cite{andras2018random, fabiani2023parsimonious}. Other works in this direction include \cite{lee2023learning,rico1994continuous,ahmadi2024ai}.\\

In the context of cancer biology, the application of physics-informed neural networks remains largely unexplored. The complexity and heterogeneity inherent in tumor biology give rise to unknown or partially characterized processes, which manifest as missing or incomplete components in mathematical models. This inherent uncertainty motivates the development of modeling frameworks capable of identifying and reconstructing missing physics or biological mechanisms from limited experimental data. Moreover, obtaining comprehensive experimental data for all the relevant components modeled remains a significant challenge due to various biological, technical, and ethical limitations, including the invasiveness of sampling, limited availability of patient tissue, variability across individuals, and the high cost and complexity of longitudinal data collection. As a result, many tumor–immune models rely on partial or noisy data, adding uncertainty to parameter estimation and model predictions.  Addressing these gaps is crucial for improving the predictive accuracy of models in oncology and for guiding the design of effective therapeutic strategies. In this work, we develop a cancer biology informed neural network framework for parameter estimation and missing physics identification. We demonstrate its robustness and generalizability using three representative ordinary differential equation (ODE) based compartmental models that exhibit distinct dynamical behaviors. The remainder of this paper is organized as follows. In Section 2, we discuss the three mathematical models used in our study. In Section 3, we discuss about structural identifiability analysis, which is a prerequisite for practical parameter estimation. Section 4 describes the method and methodology used and the architecture of CBINN. Section 5 presents the results of our experiments. Finally, Section 6 discusses the implications and potential extensions of this approach.

\section{Mathematical Models on Cancer}

As the understanding of cancer dynamics has advanced, so have the models used to describe tumor growth. Early frameworks, like the exponential growth model, assumed constant, unconstrained growth—predicting unlimited expansion. However, experimental evidence soon revealed that tumor growth slows as size increases, eventually shifting from exponential to linear behavior \cite{laird1964dynamics, spratt1995rates, spang1989growth, padmanabhan2021mathematical}. The logistic growth model addressed some of these limitations by introducing growth constraints, yet it still overlooked interactions with other cells in the tumor microenvironment. This led to the development of more sophisticated compartment-based models using nonlinear dynamical equations to better capture tumor–host interactions. In the following, we briefly describe the three compartment-based models that we consider in this study. These models are benchmark models in mathematical oncology based on which several models have come up to explore the diverse aspects of tumor dynamics. A key motivation for selecting these particular models is the distinct dynamic behaviors exhibited by these models. The first model demonstrates damped oscillations that stabilize over time, the second exhibits periodic behavior, and the third demonstrates rapid initial growth followed by stabilization at a steady-state equilibrium. These diverse behaviors allow us to evaluate the robustness and versatility of our proposed CBINN framework. By leveraging these highly nonlinear dynamics, this framework aims to infer the model parameters and uncover the underlying physical laws or mathematical structures governing these biological systems from sparse and noisy observational data. This selection of models ensures coverage of the dynamic patterns typically observed in compartmental ODE models of tumor-immune interactions, thereby validating the generalization and effectiveness of the proposed methodology.

\subsection{\textbf{Tumor-Immune Model }}

The immune response to a tumor is usually cell-mediated,
with cytotoxic T lymphocytes (CTL) and natural killer (NK) cells playing a
dominant role. The dynamics of the antitumor immune response in vivo is complicated and is not well understood. Spontaneously arising tumors are known to have low immunogenicity and usually grow outside the control of an organism. Escape from immune surveillance has been linked to a number of different
mechanisms, including the selection of tumor clones resistant to cytolytic
mechanisms, the loss or masking of tumor antigens, and tumor-induced disorders in immunoregulation \cite{brondz1988t}. However, cancer cells are attacked and killed by cells of the immune system \cite{hellstrom1969cellular}, and therefore immune surveillance of spontaneous tumors may be effective and important in keeping cancer incidence low. In \cite{kuznetsov1994nonlinear}, mathematical model of a cell-mediated
response to a growing tumor cell population is developed and the well posedness and qualitative behaviors of the system is studied. The dynamics of effector cells (x) and tumor cells (y) modeled in  \cite{kuznetsov1994nonlinear} is given by the following system of equations.
\begin{align}
\frac{dx}{dt}
&=
\sigma
\;+\;
\frac{\rho\,x\,y}{\eta + y}
\;-\;\mu\,x\,y
\;-\;\delta\,x,
\label{eq11}\\
\frac{dy}{dt}
&=
\alpha\,y\bigl(1 - \beta\,y\bigr)
\;-\;x\,y.
\label{eq21}
\end{align}
with the positive initial conditions:
\[
x(0)=x_0 > 0,\quad y(0)=y_0 > 0.
\]

 The parameter $\sigma$ is the flow rate of mature effector cells in the region of tumor cells (TC) localization, $\delta$ is a positive 
constant representing the rates of elimination of effector cells resulting from their destruction or migration from the TC localization area. The function $F(x, y) = \frac{\rho\,x\,y}{\eta + y}$ characterizes the rate at which cytotoxic effector cells 
accumulate in the region of TC localization due to the presence of the tumor. The rate of stimulated accumulation has some maximum value as $y$ gets large. This is 
consistent with limitations in the rate of transport of effector cells to the tumor.  $\mu$ is the elimination rate of the effector cells due to the presence of tumor cells. The maximal growth rate of the TC population is $\alpha$. These 
parameter incorporates both multiplication and death of TC. The maximal 
carrying capacity of the biological environment for TC (i.e. the maximum 
number of cells due, for example, to competition for resources such as oxygen, 
glucose, etc.) is $\beta^{-1}$. All the model paramters are positive and the true values are as follows.
\[
\sigma = 0.1181,\quad
\rho = 1.131,\quad
\eta = 20.19,\quad
\mu = 0.00311,\quad
\delta = 0.3743,\quad
\alpha = 1.636,\quad
\beta = 2.0\times10^{-3}.
\]

Using these chosen parameter set, we numerically solve the ODE system with the LSODA solver (SciPy's \texttt{odeint}) to generate synthetic data, which are subsequently used as observations for a neural-network inverse problem to recover the original parameters. Our objective is to develop a CBINN framework capable of accurately identifying the true model parameters and addressing the problem of gray-box identification. With "Gray-Box", we indicate the missing terms of a model. For this model $(\ref{eq11})-(\ref{eq21})$, we consider the missing term to be $\frac{\rho\,x\,y}{\eta + y}
\;-\;\mu\,x\,y$ on the right-hand side of the first ODE $(\ref{eq11})$. We approximate this term with an unknown function $f(t)$. Our goal is to show that CBINN model can accurately approximate the missing term from the available data for $x$ and $y$. 

\begin{align}
\frac{dx}{dt}
&=
\sigma
\;+\;
f(t)
\;-\;\delta\,x,
\label{eq1}\\
\frac{dy}{dt}
&=
\alpha\,y\bigl(1 - \beta\,y\bigr)
\;-\;x\,y.
\label{eq2}
\end{align}
with the positive initial conditions:
\[
x(0)=x_0 > 0,\quad y(0)=y_0 > 0.
\]

\subsection{\textbf{Tumor-Immune Model with Cytokine}}
The second model that we consider in our study is a tumor-immune model that incorporates an additional component of cytokines. Cytokines are protein hormones that mediate both natural and specific immunity. They are produced mainly by activated T cells (lymphocytes)
 during cellular-mediated immunity. Among various cytokines, interleukin-2 (IL-2) is the main cytokine responsible for lymphocyte activation, growth, and differentiation and plays an important role in tumor dynamics \cite{curti1996influence, gause1996phase, hara1996rejection}. As an extension to the model described above, in \cite{kirschner1998modeling}, the authors have explored the role of cytokines in the long-term tumor recurrence and short-term tumor oscillations. The extended model has 3 state variables and 11 parameters.  The model include: x($\tau$), the activated immune-system cells (commonly called effector cells) such as cytotoxic T-cells,
 macrophages, and natural killer cells that are cytotoxic to the tumor cells;
 y($\tau$), the tumor cells; and z ($\tau$), the concentration of IL-2. The interactions between these three components are described by the following system of non-linear equations.

\begin{align}
\frac{dx}{d\tau}
&= c\,y - \mu_{2}x + \frac{p_{1}xz}{g_{1} + z} ,
\label{eq25}\\
\frac{dy}{d\tau}
&= r_{2}\,y\bigl(1 - b\,y\bigr) - \frac{a\,x\,y}{g_{2} + y},
\label{eq26}\\
\frac{dz}{d\tau}
&= \frac{p_{2}\,x\,y}{g_{3} + y} - \mu_{3}\,z ,
\label{eq27}\\
x(0) &= x_{0}, 
\quad y(0) = y_{0}, 
\quad z(0) = z_{0}.
\label{eq28}
\end{align}

with the positive initial conditions:
\[
x(0)=x_0 > 0,\quad y(0)=y_0 > 0,\quad z(0)=z_0 > 0.
\]

Effector cells are assumed to grow based on two terms. One is a recruitment term ($cy$) due to the direct presence of the tumor, where the parameter $c$ models the antigenicity of the tumor. Antigenicity can be thought of as a measure of how
different the tumor is from self. The other growth/source term ($\frac{p_1 x y}{g_1 + z}$) is a proliferation term whereby effector cells are stimulated by IL-2 that is
 produced by effector cells. This
 term is of Michaelis-Menten form to indicate the saturated effects of the
 immune response. Effector cells have a natural lifespan of an average $\frac{1}{\mu_2}$. The tumor cells grows logistically with growth rate $r_2$ and carrying capacity $b^{-1}$. The loss of tumor cells is represented by an immune-effector cell interaction at rate $a$. This rate constant, represents the strength of the immune response
 and is modeled by MichaelisMenten kinetics to indicate the limited immune
 response to the tumor (This form could also account for the effect of a solid
 tumor, i.e. only a portion of the tumor mass comes in contact with the immune
 system cells). Equation $\ref{eq27}$ gives the rate of change for the concentration of
 IL-2. Its source is the effector cells that are stimulated by interaction with the
 tumor and also has Michaelis-Menten kinetics to account for the self-limiting
 production of IL-2. \\

For the gray-box identification problem, we denote the missing term by the unknown function $g(t)$, which comes in the last ODE (\ref{eq27}), as follows.

\begin{align}
\frac{dx}{d\tau}
&= c\,y - \mu_{2}x + \frac{p_{1}xz}{g_{1} + z} ,
\label{eq29}\\
\frac{dy}{d\tau}
&= r_{2}\,y\bigl(1 - b\,y\bigr) - \frac{a\,x\,y}{g_{2} + y},
\label{eq210}\\
\frac{dz}{d\tau}
&= g(t) - \mu_{3}\,z ,
\label{eq211}\\
x(0) &= x_{0}, 
\quad y(0) = y_{0}, 
\quad z(0) = z_{0}.
\label{eq212}
\end{align}

\subsection{\textbf{Tumor - Immune Model with Immune Cell Conversion}}

Another important aspect of tumor cells is their ability to influence immune cell phenotypes, a process known as immune cell conversion.  A
prime example of immune cell conversion is macrophage polarization between the
M1 and M2 phenotypes. In the M1 phenotype,
macrophages produce tumor suppressing molecules, including nitric oxide
and reactive oxygen species, while M2 macrophages produce pro-tumor factors and promote angiogenesis \cite{moffett2023modeling}. Studying this aspect of tumor cells in tumor-immune interaction models becomes essential, as it offers critical insights into the mechanisms of tumor growth, immune suppression, and the varying effectiveness of treatment strategies. Therefore, the third model considered in our study of parameter discovery and missing physics identification is a compartmental model that incorporates immune cell conversion. \\

In \cite{moffett2023modeling}, the authors examine a generalized Lotka-Volterra type model that describes the population densities of tumor cells (T), pro-tumor immune cells (P) and anti-tumor immune cells (A).  The inclusion of a tumor-induced switching term from anti-tumor to pro-tumor immune phenotypes makes this model unique. The model is given by the following system of non-linear equations. 

\begin{align}
\frac{dT}{dt}
&= T\Bigl(r_{T} - \tfrac{r_{T}}{K_{T}}\,T + \alpha_{TP}P - \alpha_{TA}A\Bigr),
\label{eq:T}\\
\frac{dP}{dt}
&= P\Bigl(-d_{P}P + \alpha_{PT}T\Bigr) + \omega A T,
\label{eq:P}\\
\frac{dA}{dt}
&= A\Bigl(r_{A} - \tfrac{r_{A}}{K_{A}}\,A + \alpha_{AT}T - \alpha_{AP}P\Bigr) - \omega A T.
\label{eq:A}
\end{align}

The parameters $r_{T}$ and $r_{A}$ describe the “intrinsic” growth rates of tumor and anti-tumor immune cells each in the absence of other cell types, while $K_{T}$ and $K_{A}$ denote their carrying capacities. The quadratic self‐limitation terms for tumor and anti-tumor immune cells are written as $r_{T}/K_{T}$ and $r_{A}/K_{A}$. The death rate of pro-tumor immune cells is denoted by $d_{P}$ and $\omega$ denote the anti-tumor to pro-tumor conversion parameter. The parameters $\alpha_{XY}$ represent the quadratic contributions of interactions between cell types to growth rates, where $\alpha_{XY}$ is the effect of $Y$ on the net growth rate of $X$.\\ 

For this model, we approximate the missing terms with two unknown
functions, $h(t)$ and $i(t)$, which are in the last two ODEs, as follows. We aim to identify both of these function from the available data for $T$, $P$ and $A$.

\begin{align}
\frac{dT}{dt}
&= T\Bigl(r_{T} - \tfrac{r_{T}}{K_{T}}\,T + \alpha_{TP}P - \alpha_{TA}A\Bigr),
\label{eq:T1}\\
\frac{dP}{dt}
&= h(t),
\label{eq:P1}\\
\frac{dA}{dt}
&= i(t).
\label{eq:A1}
\end{align}

\section{Structural Identifiability Analysis}
One of the fundamental challenges in designing, testing, calibrating,
 and using mathematical models comes from the fact that, in practice, often only a few of the variables are observed/measured. Despite huge amount of data and advances in technology, the appropriate data may be inaccessible. Therefore, relations involving only observable variables play an important role in systems and control theory (referred to as input-output relations; see, e.g., the textbooks and the references \cite{conte2007algebraic, wang1995orders, bellu2007daisy, ljung1994global} ).  The structural identifiability is related to the model structure independent of the experimental data and helps to address the question of unique estimation of the unknown parameters based on the postulated model. In the ODE model of the form described above in which coefficients involve unknown scalar parameters, a parameter is called \emph{structurally globally identifiable} if its value can be uniquely determined from the input–output data, assuming the absence of noise and sufficiently exciting inputs. Suppose we are given a dynamical system of the following abstract form:
\[
X' = f(X,\Theta,u), 
\quad
y = g(X,\Theta,u),
\]
where \(X=(X_{1},\dots,X_{n})\) represents the state variables, 
\(y=(y_{1},\dots,y_{m})\) represents the observables, 
\(\Theta=(\theta_{1},\dots,\theta_{k})\) contains the parameters to identify, 
and \(u\) represents the input variable to the system.  
A parameter set \(\Theta\) is called structurally \emph{globally identifiable} if
\begin{equation}\label{eq:identifiability}
  g(X,\Theta,u) = g(X,\Phi,u)
  \quad\Longrightarrow\quad
  \Theta = \Phi
  \tag{3}
\end{equation}
for every \(\Phi=(\phi_{1},\dots,\phi_{k})\) in the same space as \(\Theta\).  
Local identifiability only requires Eq.~\eqref{eq:identifiability} to hold in a neighborhood of \(\Theta\).  
As a consequence, if a model parameter turns out to be locally identifiable,  it is suggested that one should limit the search range for this parameter before fitting the model. Parameter $p$ is structurally nonidentifiable, if changing the parameter does not necessarily alter the model trajectory, because the changes can be fully compensated by altering other parameters. \\

\begin{figure}[H]
\begin{center}
\includegraphics[width=8in, height=3.3in, angle=0]{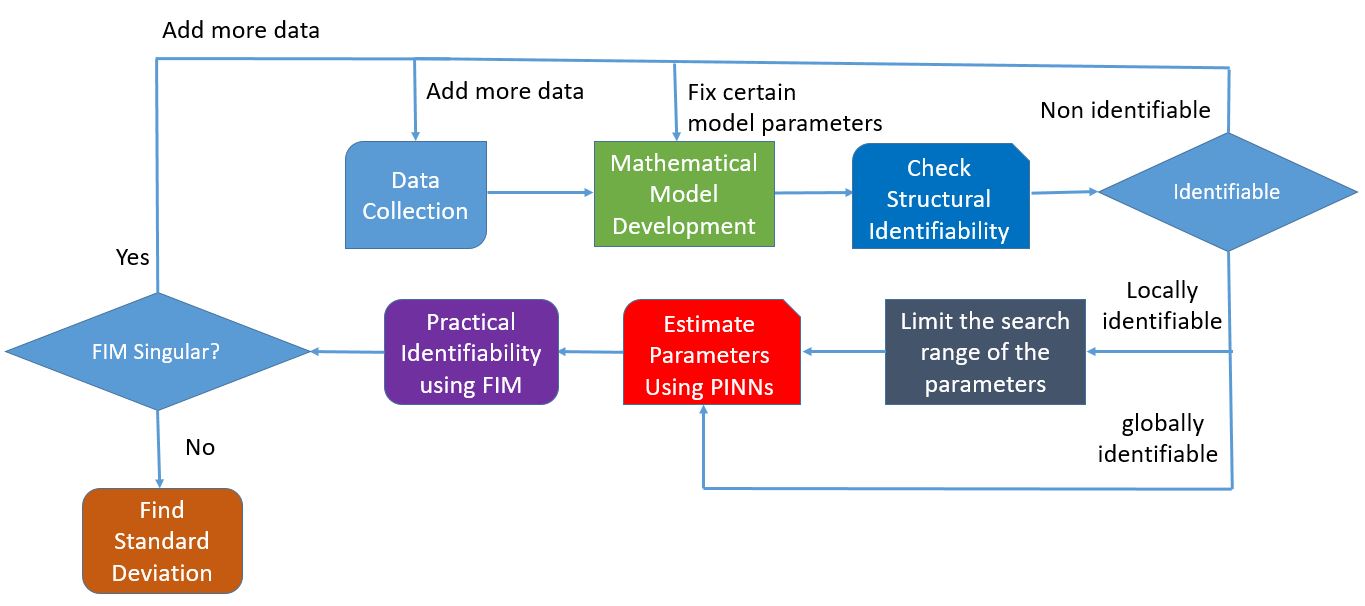}
 \caption{The workflow for the identification of model parameters.  If any parameters are not structurally identifiable, repeat data collection step to obtain additional data, or fix some parameters using prior knowledge. If the Julia StructuralIdentifiability test indicates that some parameters are only locally identifiable, then in this case we restrict the search ranges of those parameters. The parameters are estimated using the proposed CBINN and then the practical identifiability analysis is done using Fischer Information Matrix. In case of singular FIM certain parameter needs to be fixed or more data is to be collected. }
\label{structural_iden_flowchart}
\end{center}
\end{figure}

This identifiability property is a natural prerequisite for practical parameter estimation, and therefore it is an important step in the experimental design process. Insufficient data can produce very different sets of parameters without affecting the fit of data if a model is structurally nonidentifiable. To resolve the nonidentifiability issue, there are mainly two options: one is to acquire data for more species, another is to fix certain parameters as their nominal values.  Another important aspect in identifiability analysis is the practical identifiability of parameters. The practical identifiability is related to the amount and quality of experimental data. Structurally identifiable parameters may not always guarantee practically identifiability due to sparse and noisy measurements. Fisher information matrix (FIM) is generally used to find the confidence intervals of the estimated parameters and determine their practical identifiability. Practical non-identifiability is intimately related to the amount and quality of measured data and manifests in a confidence interval that is infinite.  A parameter may be structurally identifiable but practically unidentifiable if the data are sparse or noisy. The Fisher information matrix (FIM) is commonly used to assess practical identifiability. Singular FIMs or very large (effectively infinite) confidence intervals indicate parameters that are practically non-identifiable \cite{rodriguez2006hybrid}. The workflow followed for parameter discovery is given in Figure \ref{structural_iden_flowchart}. \\
 
\begin{figure}[hbt!]
\begin{center}
\includegraphics[width=8in, height=3.4in, angle=0]{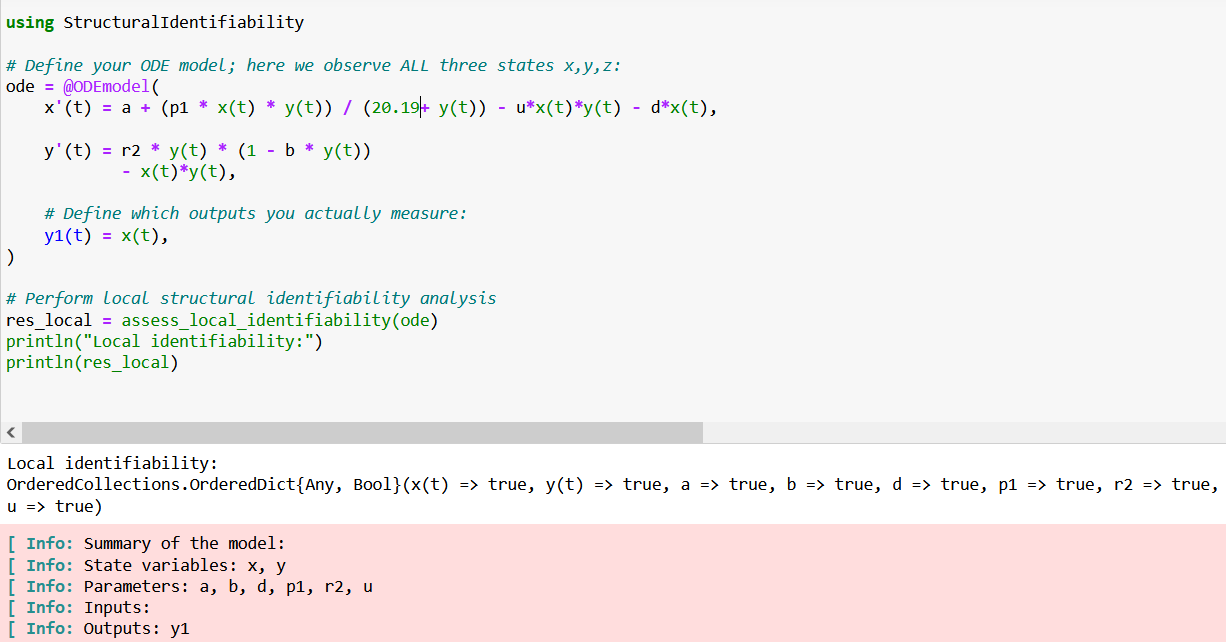}
 \caption{The output of local structural identifiability test for model (\ref{eq1}) - (\ref{eq2}) using Julia library StructuralIdentifiability \cite{dong2023differential}. We find that if $x(t)$ is observable, then fixing $\eta$, all the other parameters of the model are locally identifiable (market as true in the output of the local structural identifiability test).}
\label{fig3}
\end{center}
\end{figure}

 In this section, we test for the identifiability of the three ODE model parameters described above. To assess parameter identifiability different approaches have been proposed (see \cite{chis2011structural, hong2020global} and references therein) and several software packages and web apps have been developed  DAISY \cite{bellu2007daisy}, COMBOS \cite{meshkat2014finding}, and Structural Identifiability Toolbox \cite{ilmer2021web}. In this study, we use the Julia library StructuralIdentifiability \cite{dong2023differential} to test for structural identifiability of the model. For the first model (\ref{eq11}) - (\ref{eq21}), we assume that $x(t)$ is observable. Fixing $\eta$ to its true value, we find that the rest of the parameters are structurally identifiable (Table \ref{table_structural_identi}). To estimate $\eta$, we find that the some sparse measurements should come from both $x(t)$ and $y(t)$.  For the second model (\ref{eq25}) - (\ref{eq28}), when only \(x(t)\) is observable, fixing \(g_i\) for \(i=1,2,3\) suffices to identify all remaining parameters.  For the third model (\ref{eq:T}) - (\ref{eq:A}), every parameter is locally structurally identifiable only if all state variables are observed. If some states are unobserved, certain parameters must be fixed a prior. In particular, fixing $K_A$ to its nominal value and observing $x$ and $y$, the remaining parameters are locally structurally identifiable (Table \ref{table_structural_identi}). We will use structural identifiability results to correctly estimate the parameter values using the physics informed neural network model. The snapshot of the local structural identifiability code along with the results obtained for the model (\ref{eq11}) - (\ref{eq21}) using  Julia StructuralIdentifiability package is shown in Figure~\ref{fig3}. The practical identifiability analysis of the parameter estimates based on the FIM are discussed in the result section.

\begin{table*}[ht!]
\centering
\caption{Local structural identifiability across three models.
Ticks (\cmark) denote identifiable; crosses (\xmark) denote not identifiable; “–” indicates a fixed (not estimated) parameter, vars denote model variables.}
\label{table_structural_identi}

\setlength{\tabcolsep}{5pt}
\renewcommand{\arraystretch}{1.15}

\begin{adjustbox}{max width=\textwidth}
\begin{tabular}{@{}ll ccc cccc cccc @{}}
\toprule
 & & \multicolumn{3}{c}{\textbf{Model (\ref{eq11})--(\ref{eq21})}} &
 \multicolumn{4}{c}{\textbf{Model (\ref{eq25})--(\ref{eq28})}} &
 \multicolumn{4}{c}{\textbf{Model $(\ref{eq:T})$--$(\ref{eq:A})$}} \\
\cmidrule(lr){3-5}\cmidrule(lr){6-9}\cmidrule(lr){10-13}
\textbf{Model} & \textbf{Parameter} &
\makecell{Given\\$x(t)$} &
\makecell{Given $x(t)$\\+ fixed $\eta$} &
\makecell{Given $x(t)$\\ + $y(t)$} &
\makecell{Given\\$x(t)$} &
\makecell{Given $x(t)$\\+ fixed $g_1$} &
\makecell{Given $x(t)$\\+ fixed $g_1,g_2,g_3$} &
\makecell{All vars\\observable} &
\makecell{Given\\$T(t)$} &
\makecell{Given\\$P(t), A(t)$} &
\makecell{Given\\$T(t), P(t)$\\ + fixed $K_A$} &
\makecell{All vars\\observable} \\
\midrule
(\ref{eq1})--(\ref{eq2}) & $\sigma$ & \cmark & \cmark & \cmark &  &  &  &  &  &  &  &  \\
 & $\rho$   & \cmark & \cmark & \cmark &  &  &  &  &  &  &  &  \\
 & $\eta$   & \xmark & –      & \cmark &  &  &  &  &  &  &  &  \\
 & $\mu$    & \xmark & \cmark & \cmark &  &  &  &  &  &  &  &  \\
 & $\delta$ & \cmark & \cmark & \cmark &  &  &  &  &  &  &  &  \\
 & $\alpha$ & \cmark & \cmark & \cmark &  &  &  &  &  &  &  &  \\
 & $\beta$  & \xmark & \cmark & \cmark &  &  &  &  &  &  &  &  \\
\midrule
(\ref{eq25})--(\ref{eq28}) & $c$     &  &  &  & \xmark & \xmark & \cmark & \cmark &  &  &  &  \\
 & $\mu_2$ &  &  &  & \cmark & \cmark & \cmark & \cmark &  &  &  &  \\
 & $p_1$   &  &  &  & \cmark & \cmark & \cmark & \cmark &  &  &  &  \\
 & $g_1$   &  &  &  &  \xmark    & –      & - & \cmark &  &  &  &  \\
 & $r_2$   &  &  &  & \xmark & \xmark & \cmark & \cmark  &  &  &  &  \\
 & $b$     &  &  &  & \xmark & \xmark & \cmark & \cmark  &  &  &  &  \\
 & $a$     &  &  &  & \xmark & \xmark & \cmark & \cmark  &  &  &  &  \\
 & $g_2$   &  &  &  & \xmark &  \xmark      & - & \cmark  &  &  &  &  \\
 & $p_2$   &  &  &  & \cmark & \cmark & \cmark & \cmark &  &  &  &  \\
 & $g_3$   &  &  &  & \xmark &  \xmark     & - & \cmark  &  &  &  &  \\
 & $\mu_3$ &  &  &  & \cmark & \cmark & \cmark & \cmark &  &  &  &  \\
\midrule
$(\ref{eq:T})$--$(\ref{eq:A})$ & $r_T$          &  &  &  &  &  &  &  & \cmark & \cmark & \cmark & \cmark \\
 & $K_T$        &  &  &  &  &  &  &  & \cmark & \xmark & \cmark & \cmark \\
 & $d_P$        &  &  &  &  &  &  &  & \xmark & \cmark & \cmark & \cmark \\
 & $r_A$        &  &  &  &  &  &  &  & \cmark & \cmark & \cmark & \cmark \\
 & $K_A$        &  &  &  &  &  &  &  & \xmark & \cmark & - & \cmark \\
 & $\alpha_{TP}$&  &  &  &  &  &  &  & \xmark & \cmark & \cmark & \cmark \\
 & $\alpha_{TA}$&  &  &  &  &  &  &  & \xmark & \cmark & \cmark & \cmark \\
 & $\alpha_{PT}$&  &  &  &  &  &  &  & \cmark & \xmark & \cmark & \cmark \\
 & $\alpha_{AT}$&  &  &  &  &  &  &  & \xmark & \xmark & \cmark & \cmark \\
 & $\alpha_{AP}$&  &  &  &  &  &  &  & \xmark & \cmark & \cmark & \cmark \\
 & $\omega$     &  &  &  &  &  &  &  & \xmark & \xmark & \cmark & \cmark \\
\bottomrule
\end{tabular} 
\end{adjustbox}
\end{table*}

\vspace{2cm}
\section{CBINN Method and Methodology}

In this section, we present the details of the CBINN model used for unknown parameter identification and gray-box identification problems.

\subsection{\textbf{CBINN for Tumor-Immune Model}}
 By incorporating the governing equations directly into the loss function, our CBINN learns the solutions that are consistent with both observed data and the fundamental physics of the system.  The tumor-immune model (\ref{eq1})-(\ref{eq2}) is used to test the predictive performance of the proposed CBINN model for two cases: 1) the unknown parameter discovery, and 2) missing physics discovery.  The CBINN network characterized by parameters $\theta$ takes time $t$ as input and generates an output vector $\hat{u}(t, \theta) = (\hat{u}_1(t, \theta), \hat{u}_2(t, \theta))$ that is an approximation of the true ODE solution. This output of the network must satisfy the data observations as well as the ODE system. For this, we have to construct the loss function having terms related to the given data observations and the ODE system. Let us assume that we have the measurement data of $u(t) = (u_1(t), u_2(t))$ at the time instances $t_1, t_2, \ldots, t_{S_\text{data}}$. Let $\tau_1, \tau_2, \ldots, \tau_{N_\text{ode}}$ be the collocation points where we enforce the network to satisfy the ODE system. Here $N$ is the total number of collocation points, and $S$ is the total number of data points. Then, the total loss function for the case of parameter discovery is defined as a function of both the neural network parameter (${\theta}$) and the parameters ${p}$ of the governing system of equations.
\begin{equation}
   \text{Total Loss} = \mathcal{L}({\theta}, {p}) = {Loss}_{\text{Data}}({\theta}) + {Loss}_{\text{ODE}}({\theta}, {p}) + {Loss}_{\text{Aux}}({\theta}) \label{loss}
\end{equation}

Here ${Loss}_{\text{Data}}$ is the mean square loss between the real data points and the model-predicted output at $t_1, t_2, \ldots, t_{N_\text{data}}$. ${Loss}_{\text{ODE}}$ is the residual loss at the collocation points. $\mathcal{L}_{\text{Aux}}$ is an additional constraint introduced for the
parameter identification of the system and involves two time instants $T_0$ and $T_1$. $T_1$ could be an arbitrary time instant within the training time window (not too close to $T_0$). In this work, we consider $T_1$ as the midpoint, and this is in line with the choice made by the authors in \cite{yazdani2020systems}. This loss is different from the data loss, because $\mathcal{L}_{\text{data}}$ is only with respect to the variables that are observable, whereas in the auxiliary loss data are given for all state variables at these two time instants. Inclusion of $\mathcal{L}_{\text{Aux}}$ in model training has been found to be very effective, especially in predicting the dynamics of nonobserved species \cite{yazdani2020systems}.   Each term of the loss function (\ref{loss}) is defined as follows.

\begin{equation}
    \text{Loss}_{\text{Data}}({\theta}) = \sum_{m=1}^{2} w^{\text{data}}_m \mathcal{L}^{\text{data}}_m = \sum_{m=1}^{2} w^{\text{data}}_m \left[ \frac{1}{N_\text{data}} \sum_{n=1}^{N_\text{data}} \left( u_m(t_n) - \hat{u}_m(t_n; \boldsymbol{\theta}) \right)^2 \right]  \label{dataloss}, 
\end{equation}

\begin{equation}
\begin{split}
\text{Loss}_{\text{ODE}}({\theta}, {p}) &= 
w_1^{\text{ode}} \frac{1}{N_{\text{ode}}} \sum_{n=1}^{N_{\text{ode}}}
\left( \left. \frac{d \hat{u}}{dt} \right|_{\tau_n} 
- \left[ \sigma + \frac{\rho \hat{x}(\tau_n) \hat{y}(\tau_n)}{\eta + \hat{y}(\tau_n)} 
- \mu \hat{x}(\tau_n) \hat{y}(\tau_n) - \delta \hat{x}(\tau_n) \right] \right)^2 \\
&+ w_2^{\text{ode}}  \frac{1}{N_{\text{ode}}} \sum_{n=1}^{N_{\text{ode}}}
\left( \left. \frac{d \hat{y}}{dt} \right|_{\tau_n} 
- \left[ \alpha \hat{y}(\tau_n)(1 - \beta \hat{y}(\tau_n)) - \hat{x}(\tau_n) \hat{y}(\tau_n) \right] \right)^2
\end{split}  \label{ode_loss}
\end{equation}

\begin{equation}
    \text{Loss}_{\text{Aux}}({\theta}) = \sum_{s=1}^{2} w^{\text{aux}}_s \mathcal{L}^{\text{aux}}_s = \sum_{s=1}^{2} w^{\text{aux}}_s \left[ \frac{(u_s(T_0) - \hat{u}_s(T_0; {\theta}))^2 + (u_s(T_1) - \hat{u}_s(T_1; {\theta}))^2}{2} \right] \label{auxloss}.
\end{equation}

We have in total six coefficients
$w_1^{\text{data}}$, $w_2^{\text{data}}$ in data loss (\ref{dataloss}), 
$w_1^{\text{ode}}, w_2^{\text{ode}}$ in (\ref{ode_loss}), and 
$w_1^{\text{aux}}, w_2^{\text{aux}}$ in  (\ref{auxloss}). In this study, we manually select these weight coefficients so that the weighted losses are of the same order of magnitude during network training. We note that this guideline makes weight selection much easier,  although there are many weights to be determined.  In network training, we have both supervised and unsupervised losses.  $\mathcal{L}_{\text{IC}}, \mathcal{L}_{\text{data}}$ and $\mathcal{L}_{\text{Aux}}$ represent the discrepancies between the predictions of the neural network and the measured data, which makes them supervised losses. Conversely, $\mathcal{L}_{\text{ode}}$ is the residual error derived
from the ODE system and, therefore, this is an unsupervised loss. We determine the parameters of the neural network $\theta^*$ and the parameters of the ODE model $p^*$ by minimizing the loss function using gradient-based optimization methods, such as the Adam optimizer \cite{kinga2015method}. These parameters are updated in each epoch by solving the following optimization problem. 
\[
\theta^*,  p^*
= \arg \min_{\theta, p} 
\mathcal{L}(\theta, p)
\] \\

For the gray-box identification problem, we have to have two neural network models. The first network characterized by parameters $\theta_1$ takes time $t$ as input and generates an output vector $\hat{u}(t, \theta_1) = (\hat{u}_1(t, \theta_1), \hat{u}_2(t, \theta_1))$ which is an approximation of the true ODE solution ${u}(t)$. The second network,  with a different design, approximates the unknown term $f(t)$. This network is characterized by parameters $\theta_2$,  takes time $t$ as input, and generates an output function $f(t, \theta_2)$. Here we assume that few noisy measurements are available for both model variables $x(t)$ and $y(t)$.  The total loss function to be minimized during the model training is computed as follows. 

\begin{equation}
   \mathcal{L}({\theta_1, \theta_2}) = \mathcal{L}_{\text{data}}({\theta_1}) + \mathcal{L}_{\text{IC}}({\theta_1})+ \mathcal{L}_{\text{ode}}({\theta_1}, \theta_2) \label{loss2}
\end{equation}

$\mathcal{L}_{\text{IC}}(\theta_1)$ and $\mathcal{L}_{\text{data}}(\theta_1)$ are the initial condition loss and data loss. $\mathcal{L}_{\text{ode}}(\theta_1, \theta_2)$ is the residual loss that is defined as
\begin{align*}
\mathcal{L}_{\text{ode}}(\theta_1, \theta_2) &= \frac{1}{N_{\text{ode}}} \sum_{n=1}^{N_{\text{ode}}} \left( 
\left. \frac{d\hat{u}}{dt} \right|_{\tau_n} - F(\tau_n, \hat{u}(\tau_n; \theta_1), f(\tau_n; \theta_2))
\right)^2, \\
\end{align*}

where $$F(\tau_n, u(\tau_n; \theta_1), f(\tau_n; \theta_2)) = \bigg( \sigma + f(\tau_n; \theta_2)  - \delta \hat{u_x}(\tau_n), \hspace{.2cm} \alpha \hat{u_y}(\tau_n)(1 - \beta \hat{u_y}(\tau_n)) - \hat{u_x}(\tau_n) \hat{u_y}(\tau_n) \bigg)$$.

We determine both neural network parameters $\theta_1^*, \theta_2^*$ by minimizing the loss function. The network parameters are updated in each epoch by solving the following optimization problem. 
\[
\theta_1^*, \theta_2^*
= \arg \min_{\theta_1, \theta_2} 
\mathcal{L}(\theta_1, \theta_2)
\]

Our goal here is to first estimate the model parameters and then identify the missing term $f(t, \theta_2)$ from the given data points. Following the motivation from \cite{yazdani2020systems}, to expedite neural network training and correctly capture model dynamics,  we introduce an additional layer in CBINN network as follows: \\

\begin{figure}[hbt!]
\begin{center}
\includegraphics[width=8in, height=4.0in, angle=0]{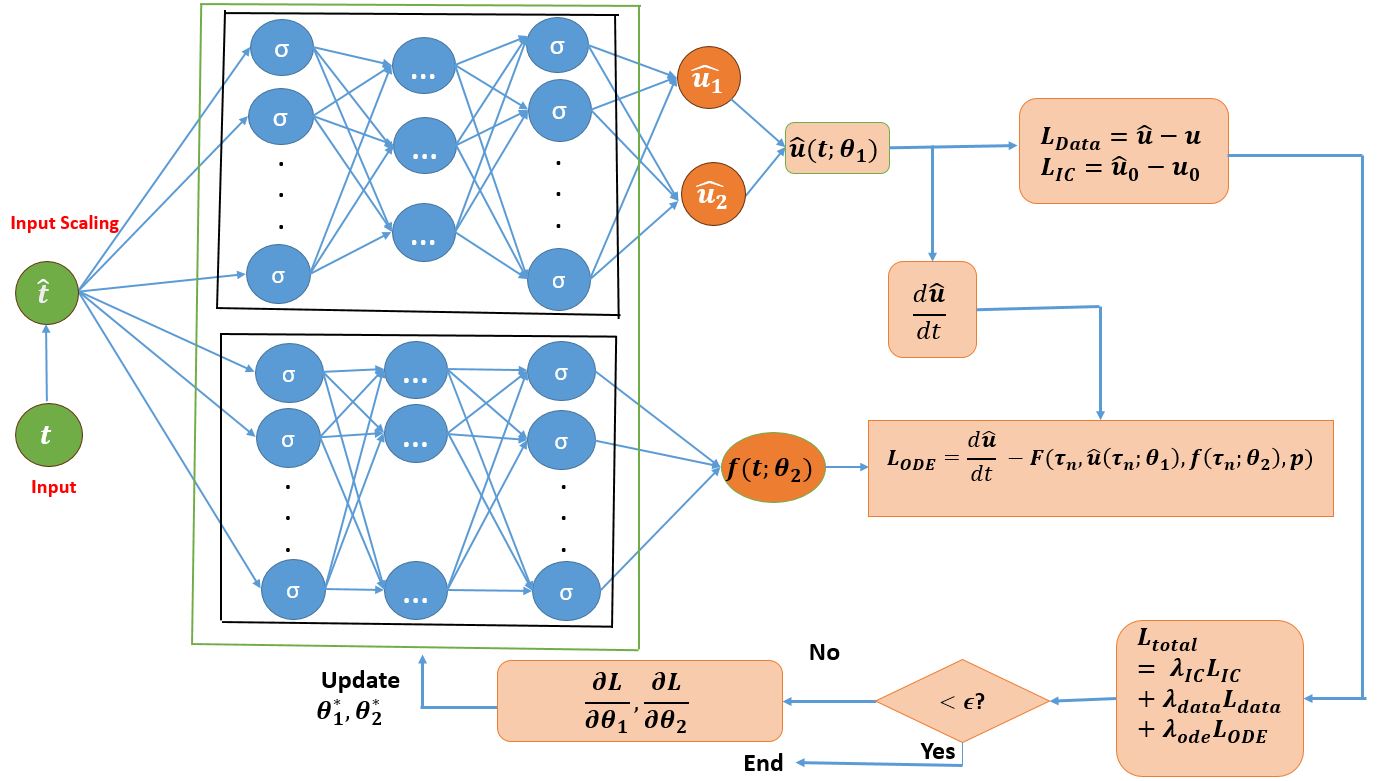}
 \caption{ Schematic of the CBINN algorithm for discovering unknown term $f(t; \theta_2)$. The scaled input \(\hat{t}\) is fed to two fully-connected sub-networks, a solution network \(\hat{u}(t;\theta_1)\) (top) that predicts the states and a dynamics network \(f(t;\theta_2)\) (bottom) that predicts the missing terms  each composed of several hidden layers with nonlinear activation function \(\sigma\).  The time derivative \(d\hat{u}/dt\) is computed by automatic differentiation and combined with \(f\) to form an ODE residual, and training minimizes the weighted sum of the data, initial-condition and ODE residual losses to estimate \(\theta_1,\theta_2\). The loss terms are defined as mean squared error (MSE) losses.}
\label{pinns_f}
\end{center}
\end{figure}

Input Scaling Layer: The input time $t$ could vary by orders of magnitude. Therefore, for robust and effective training, we introduce the input scaling layer. We divide $t$ by its maximum value so that $ t \sim \mathcal{O}(1)$. We define $$\tilde t = \frac{t}{T},
      \qquad\text{such that}\quad \tilde t \sim \mathcal{O}(1).$$
and use $\tilde t$ instead of $t$ for model training. $T$ is the maximum value of the time domain. The schematic of the CBINN algorithm for the discovery of the unknown term $f(t; \theta_2)$ for the tumor immune model is shown in Figure \ref{pinns_f}. For parameter estimation problem, we have just the first main network. The method and methodology for the second model (tumor-immune model with cytokine) are almost the same as the one we just described, but for the third model ((\ref{eq:T}) - \ref{eq:A}) it changes slightly due to two unknown functions. We briefly discuss the details of the third model in the following.

\subsection{\textbf{CBINN for Tumor - Immune Model with Immune Cell Conversion}}

The model $(\ref{eq:T})$--$(\ref{eq:A})$ has eleven parameters to estimate. The structural identifiability analysis indicates that, if the parameter \(K_A\) is fixed prior to its nominal value, the remaining model parameters can be structurally identifiable with measurements of $T$ and $P$. For the gray-box identification problem, the governing system of ODEs for this model is as
follows:

\begin{align}
\frac{dT}{dt}
&= T\Bigl(r_{T} - \tfrac{r_{T}}{K_{T}}\,T + \alpha_{TP}P - \alpha_{TA}A\Bigr),
\label{eq:T11}\\
\frac{dP}{dt}
&= h(t, \theta_2),
\label{eq:P11}\\
\frac{dA}{dt}
&= i(t, \theta_2).
\label{eq:A11}
\end{align}

\begin{figure}[hbt!]
\begin{center}
\includegraphics[width=8in, height=4.0in, angle=0]{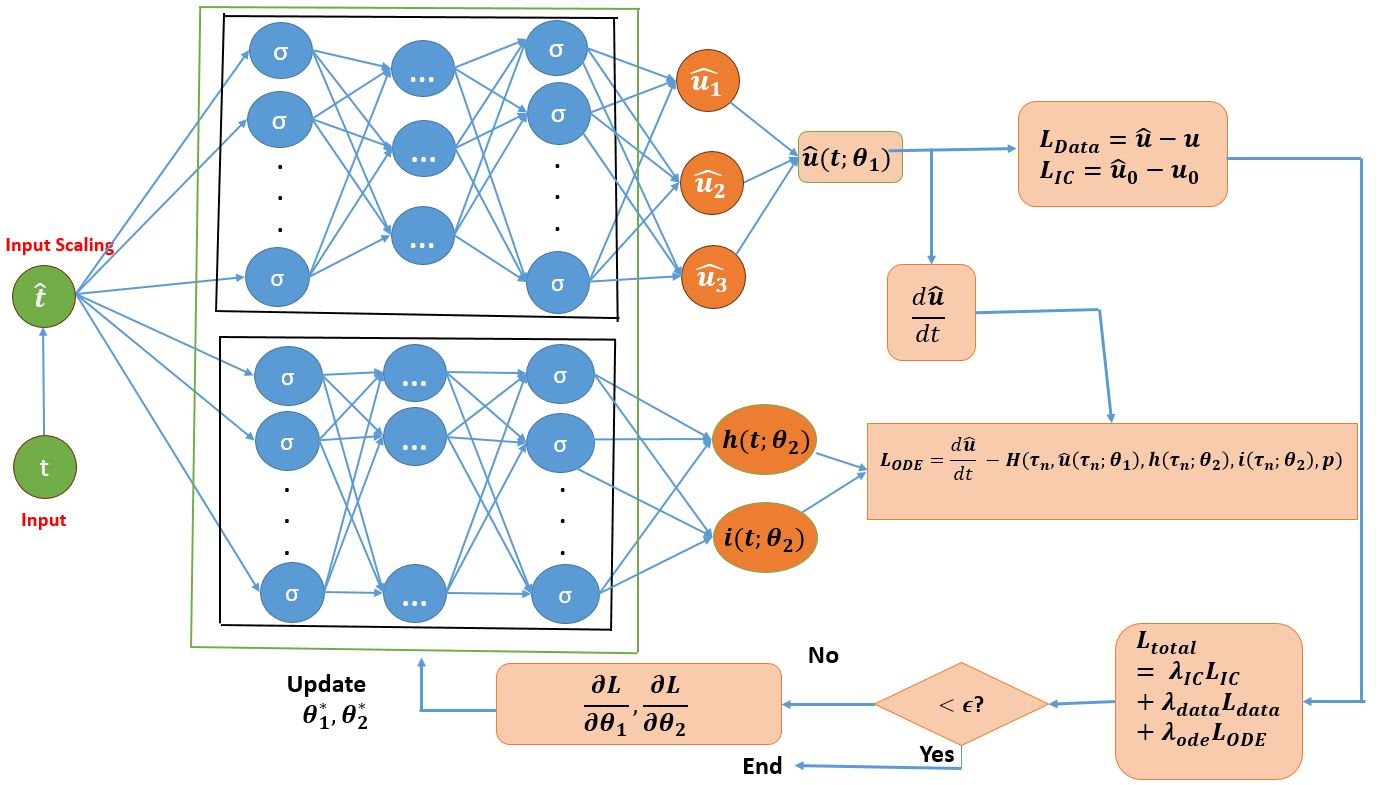}
 \caption{ Schematic of the CBINN algorithm for predicting the unknown term $h(t; \theta_2)$ and $i(t; \theta_2)$.  We have two networks, one to approximate the ODE solution and the other to approximate the unknown functions.}
\label{pinns_archi_model3}
\end{center}
\end{figure}

In this setup, the first network, parameterized by $\theta_1$, takes time as input and outputs the approximate state trajectory $\hat{ u}(t;\theta_1)$. The second network characterize by $\theta_2$ takes time as an input and generates
two outputs $h(t; \theta_2)$ and $i(t; \theta_2)$. Using automatic differentiation, we compute $\mathrm{d}\hat{ u}/\mathrm{d}t$ and form the ODE residual by combining $\hat{ u}$, $h$, $i$, and the trainable model parameters. The total loss comprising data and initial-condition terms together with the ODE residual is minimized by backpropagation to jointly learn the trainable parameters.  A schematic of the algorithm for this case is shown in Figure \ref{pinns_archi_model3}.

\section{Results}
In this section, we present and discuss the results obtained from our simulations. The performance of the CBINN model in parameter estimation and gray-box identification is evaluated using synthetic data generated by numerically solving the three compartmental models described earlier. To better represent the realistic scenarios and verify the robustness of the algorithm against noise, the synthetic data are perturbed with additive Gaussian noise at varying levels.

\subsection{\textbf{Tumor-Immune Model}}

First, we aim to infer the value of the parameters $\sigma, \rho, \eta , \mu, \delta , \alpha , \beta $
of the ODE system (\ref{eq11}) - (\ref{eq21}) and the hidden dynamics of the unobserved variables of the model given sparse noisy observational data of the observed variable. The target values of the parameters are as follows.
\[
\sigma = 0.1181,\quad
\rho = 1.131,\quad
\eta = 20.19,\quad
\mu = 0.00311,\quad
\delta = 0.3743,\quad
\alpha = 1.636,\quad
\beta = 2.0\times10^{-3}.
\]

\begin{figure}[H]
\begin{center}
\includegraphics[width=8in, height=3in, angle=0]{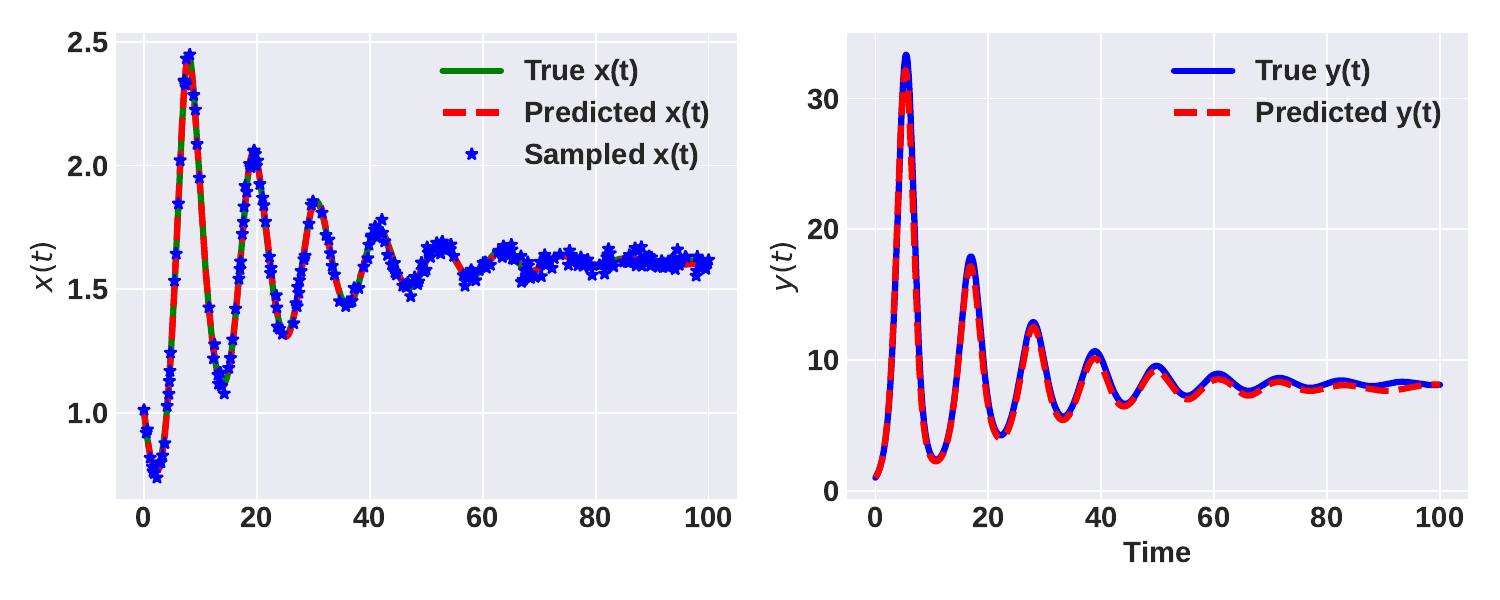}
 \caption{ CBINN inferred dynamics of tumor-immune model $(\ref{eq11})$--$(\ref{eq21})$ from noisy observations compared with the true solution. True solutions are generated by solving the system (\ref{eq11})-(\ref{eq21}) using the LSODA algorithm implemented in Python’s scipy.integrate.odeint function with the target parameter values. Predictions are
 performed on equally-spaced time instants in the interval of $0 - 100$ days. 200 measurements are corrupted by a zero-mean Gaussian noise and standard deviation of $\sigma^* = 0.05c $ where $c$ is the standard deviation of the
 observable variable ($x$) over the observation time window.  }
\label{fig2model1}
\end{center}
\end{figure}

Based on the structural identifiability analysis (Table \ref{table_structural_identi}), we consider two observation scenarios. Case 1: $x(t)$ is the only observable variable, fix $\eta$ to its true value, and estimate the rest of the parameters. In case 2, we assume that some measured data are available for both $x(t)$ and $y(t)$, and we use CBINN to estimate the full set of parameters. Synthetic data are generated by solving the system (\ref{eq11})-(\ref{eq21}) using the LSODA algorithm implemented in Python’s scipy.integrate.odeint function with the target parameter values.  The simulation was carried out over the time interval t = 0 to t = 100 (days), with a total of 500 equally spaced time points and initial condition $(1.0, 1.0)$. The numerical solution of the system obtained using the LSODA solver is shown in Figure \ref{fig2model1}.  The synthetic data generated are incorporated into the data loss part of the total loss that is optimized. To represent experimental noise, we corrupt the observation data by Gaussian noise with zero mean and standard deviation $\sigma^* = c s$ where $c$ is the standard deviation of each observable variable over the observation time window and $s = $ 0 to 0.1. We evaluate the robustness of the model for three different values of $s$ - $0.02, 0.05,  0.1$. The hyperparameters of the neural network model are shown in Table \ref{table_pinns_hyperparamter}. \\

\begin{figure}[hbt!]
\begin{center}
\includegraphics[width=8in, height=4.5in, angle=0]{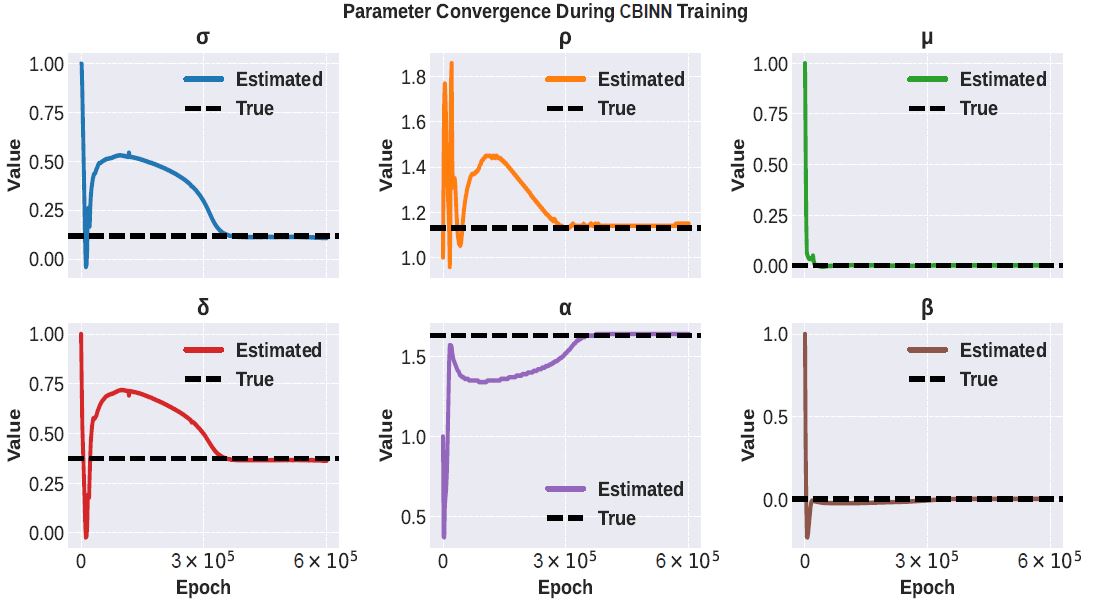}
 \caption{ Parameter convergence during CBINN training for model (\ref{eq11})-(\ref{eq21}). Solid curves show the estimated parameter values across epochs; black dashed lines indicate the ground truth values. All parameters converge to their targets, with $\mu$ and $\beta$ stabilizing rapidly, $\rho$ converging after an oscillatory transient, and $\sigma$, $\delta$, and $\alpha$ exhibiting slower but steady approaches. }
\label{convergence_para_model1}
\end{center}
\end{figure}

\begin{figure}[hbt!]
\begin{center}
\includegraphics[width=7.5in, height=3.2in, angle=0]{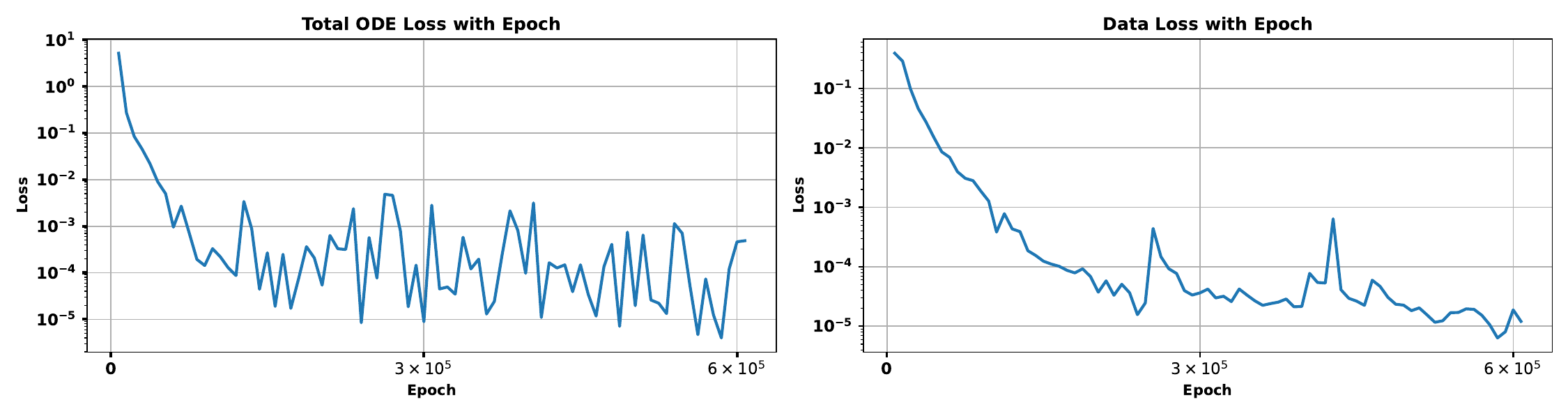}
 \caption{Total ODE residual loss (sum of the two equation residuals) and data loss (right) plotted against iterations with swish activation function when $x(t)$ is observable for the model (\ref{eq11})-(\ref{eq21}).  }
\label{loss_fn_model1}
\end{center}
\end{figure}

\begin{figure}[hbt!]
\begin{center}
\includegraphics[width=7.5in, height=3.7in, angle=0]{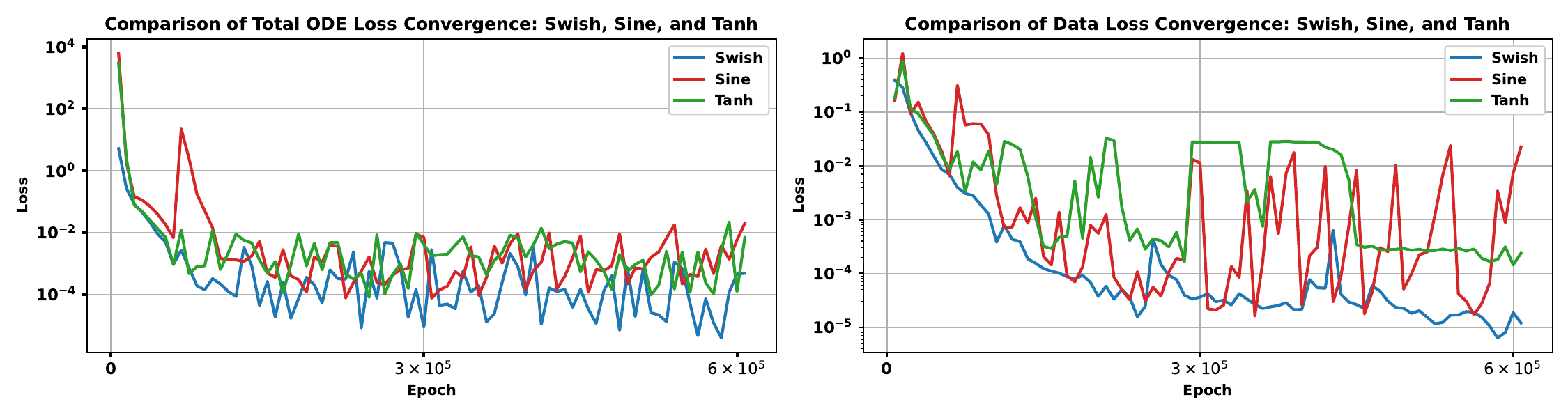}
 \caption{ODE and data loss comparisons for Swish, Tanh and Sine activations for model (\ref{eq11})-(\ref{eq21}). Swish achieves the lowest and most stable losses; Tanh is moderate; Sine exhibits large oscillations. }
\label{loss_compare_model1}
\end{center}
\end{figure}

For case 1, 200 noiseless measurements are sampled at random from the generated synthetic data. Since in this case the measured data are available only for $x(t)$, the data loss is only with respect to $x(t)$. For residual error, we sample the inputs (collocation points) within the time frame of $0-100$ at random. For neural network training, we utilized Adam's optimization with 5000 collocation points, swish activation function $(S(x) = x \cdot \sigma(x) = \frac{x}{1 + e^{-x}})$, learning rate $0.001$, and trained the model for 600,000 iterations.  The results of the CBINN inferred dynamics and the true solution for the noise level of $5\%$ is shown in Figure \ref{fig2model1}. We observe excellent agreement between the inferred and exact dynamics. The neural network model learns the input data and is able to correctly infer the dynamics of the unobserved variable $y$ due to the constraints imposed by the ODE system. We executed the model five times, each with a distinct random seed, and calculated the mean absolute error between the inferred parameter values and the target values. The target values, mean inferred values, and mean absolute errors with and without noise in the observations are given in Table \ref{table1_model1}. The agreement between the inferred and exact dynamics is excellent considering the relatively high level of noise in the training data.  The estimated parameter trajectories with iterations during training against their ground truth values are shown in Figure \ref{convergence_para_model1}. All six parameters approach their true values, demonstrating stable and well-posed identification. Figure \ref{loss_fn_model1} shows the total ODE and data losses with the swish activation function. Figure \ref{loss_compare_model1} compares losses with respect to swish, sine and tanh activations. The losses with respect to swish activation is the lowest in this case and shows most stable convergence. In case 2 we collected 100 measurements (with and without noise) at random from each variable $x$ and $y$; as in case 1, the CBINN-inferred dynamics show excellent agreement with the exact solution. The inferred parameter values and the errors are shown in Table \ref{table2_model1}. \\

\begin{table}[h!]
\centering
\captionsetup{justification=centering}
\caption{Hyper-parameters of the CBINN Network. For the parameter-estimation problem we train only the main network; all other settings (activation function, optimizer, learning-rate schedule) remains the same. In the Number of Iterations (epochs) row the first value is used for parameter discovery and the second for gray-box identification.}
\renewcommand{\arraystretch}{1.2}
\begin{tabular}{|p{6cm}|c|c|c|}
\hline
\textbf{CBINN parameters} & \textbf{Model 1 $(\ref{eq11})$--$(\ref{eq21})$ } & \textbf{Model 2 $(\ref{eq25})$--$(\ref{eq28})$} & \textbf{Model 3 $(\ref{eq:T})$--$(\ref{eq:A})$} \\
\hline
Optimizer & Adam & Adam & Adam \\
\hline
Activation Function & Swish & Swish & Swish \\
\hline
Number of Iterations (epochs) & $6 \times 10^5$, $2 \times 10^5$ & $9 \times 10^5$, $2 \times 10^5$ & $5 \times 10^5$, $1 \times 10^5$ \\
\hline
Width and Depth of main NN & 32, 6 & 32, 7 & 64, 5 \\
\hline
Width and Depth of second NN & 32, 4 & 32, 4 & 32, 4 \\
\hline
Learning Rate for main NN & 0.001 & 0.001 & 0.001 \\
\hline
Learning Rate for second NN & 0.001 & 0.001 & 0.001 \\
\hline
Number of Collocation Points & 5000 & 10000 & 5000 \\
\hline
\end{tabular} \label{table_pinns_hyperparamter}
\end{table}

\begin{table}[tbp]
  \centering
  \caption{Case 1 (given $x(t)$ as observable variable and fixed $\eta$): Parameter estimates and mean absolute errors for the tumor–immune model (\ref{eq11})-(\ref{eq21}) under different noise levels. Standard deviation for each parameter is calculated using Fischer information matrix.}
  \label{table1_model1}
  \setlength{\tabcolsep}{4pt}
  \small
  \begin{tabular}{l
                  S[table-format=1.5]     
                  *{4}{S[table-format=1.5]}
                  *{4}{S[table-format=1.5]}
                  S[table-format=1.5]     
                  }
    \toprule
    & & \multicolumn{4}{c}{\textbf{Estimated value}} & \multicolumn{4}{c}{\textbf{Mean Absolute error}} & \multicolumn{1}{c}{\textbf{Std.\ dev.}} \\
    \cmidrule(lr){3-6}\cmidrule(lr){7-10}\cmidrule(lr){11-11}
    \textbf{Parameter} & \textbf{Target}
      & \textbf{Noiseless} & \textbf{2\%} & \textbf{5\%} & \textbf{10\%}
      & \textbf{Noiseless} & \textbf{2\%} & \textbf{5\%} & \textbf{10\%}
      & \textbf{} \\
    \midrule
    $\sigma$ & 0.1180 & 0.1080 & 0.1062 & 0.0990 & 0.0973 & 0.0101 & 0.0117 & 0.0191 & 0.0206 & 0.0072\\
    $\rho$   & 1.1300 & 1.0812 & 1.1012 & 1.1221 & 1.1320 & 0.0487 & 0.0288 & 0.0079 & 0.0020 & 0.0230 \\
    $\mu$    & 0.0031 & 0.0028 & 0.0025 & 0.00275 & 0.00264 & 0.0003 & 0.0006 & 0.00035 & 0.00046 & 0.0004 \\
    $\delta$ & 0.3743 & 0.3641 & 0.3522 & 0.3587 & 0.3484 & 0.0102 & 0.0221 & 0.0156 & 0.0258 & 0.0055 \\
    $\alpha$ & 1.6360 & 1.6400 & 1.6221 & 1.6510 & 1.6620 & 0.0041 & 0.0139 & 0.0150 & 0.0261 & 0.0045 \\
    $\beta$  & 0.0020 & 0.0024 & 0.0022 & 0.0025 & 0.0032 & 0.0004 & 0.0002 & 0.0005 & 0.0013 & 0.0003 \\
    \bottomrule
  \end{tabular}
\end{table}


\begin{table}[tbp]
  \centering
  \caption{Case 2 (both $x(t)$ and $y(t)$ are observables): Parameter estimates and mean absolute errors for the tumor–immune model (\ref{eq11})–(\ref{eq21}) under different noise levels.}
  \label{table2_model1}
  \setlength{\tabcolsep}{4pt}
  \small
  \begin{tabular}{l
                  S[table-format=2.5]
                  *{4}{S[table-format=2.5]}
                  *{4}{S[table-format=1.5]}}
    \toprule
    & & \multicolumn{4}{c}{\textbf{Estimated value}} & \multicolumn{4}{c}{\textbf{Mean Absolute error}} \\
    \cmidrule(lr){3-6}\cmidrule(lr){7-10}
    \textbf{Parameter} & \textbf{Target}
      & \textbf{Noiseless} & \textbf{2\%} & \textbf{5\%} & \textbf{10\%}
      & \textbf{Noiseless} & \textbf{2\%} & \textbf{5\%} & \textbf{10\%} \\
    \midrule
    $\sigma$ & 0.1180 & 0.1160 & 0.1091 & 0.0990 & 0.0983 & 0.0021  & 0.0088  & 0.0190  & 0.0196 \\
    $\rho$   & 1.1300 & 1.1120 & 1.1013 & 1.1310 & 1.1500 & 0.0180  & 0.0287  & 0.0010  & 0.0200 \\
    $\eta$   & 20.190 & 20.090 & 19.980 & 19.950 & 19.920 & 0.1000  & 0.2100  & 0.2400  & 0.2701 \\
    $\mu$    & 0.0031 & 0.0030 & 0.0028 & 0.0027 & 0.00265 & 0.00010 & 0.00030 & 0.00040 & 0.00045 \\
    $\delta$ & 0.3743 & 0.3750 & 0.3670 & 0.3620 & 0.3482 & 0.00070 & 0.00730 & 0.01231 & 0.02610 \\
    $\alpha$ & 1.6360 & 1.6400 & 1.6250 & 1.6310 & 1.6520 & 0.0040  & 0.0110  & 0.0050  & 0.0160 \\
    $\beta$  & 0.0020 & 0.0019 & 0.0021 & 0.0023 & 0.0032 & 0.00010 & 0.00010 & 0.00030 & 0.00120 \\
    \bottomrule
  \end{tabular}
\end{table}


\vspace{0.5cm}

Now we find the standard deviation of the parameter estimates and determine their practical identifiability, using Fisher information matrix (FIM). For a dynamic ODE model with state vector \(x(t)\) and parameter vector \(p\), the FIM is constructed by summing the sensitivity contributions at the measurement times:
\begin{equation}
\label{eq:FIM}
F \;=\; \sum_{t_n} S(t_n)^{\!\top}\, C^{-1}\, S(t_n),
\end{equation}
where \(C\) is the measurement error covariance matrix and
$ S(t)  =  \frac{\partial x(t)}{\partial p} $
is the (time-dependent) sensitivity matrix.  The standard error for the parameter \(p_i\) is approximated by
$\sigma_i \approx \sqrt{(F^{-1})_{ii}},
$ and the 95$\%$ confidence intervals are given by $p_i \pm 2\,\sigma_i$. A (nearly) singular FIM indicates parameters that are practically nonidentifiable given the experiment (assuming structural identifiability holds). The high correlations among parameters may lead to a singular FIM. Large absolute correlations (close to 1) between the parameters suggest strong dependencies and weak practical identifiability.  We check the eigenvalues and compute the FIM eigenvectors associated with the zero eigenvalues (i.e., null eigenvectors) to look for nonidentifiable parameters. A null eigenvector that has a dominant component associated with a single parameter indicates that changes in this parameter do not affect the state variables.
We also construct the correlation matrix R given by $R_{ij} \;=\; \frac{(F^{-1})_{ij}}{\sqrt{(F^{-1})_{ii}\,(F^{-1})_{jj}}} $ and search for correlations between the parameters. More details on the Fisher information matrix (FIM) and the corelation matrix are given in \cite{forman2012genetic} and in the supplementary material of \cite{yazdani2020systems}.
 \\

\begin{figure}[hbt!]
  \centering
  \begin{subfigure}[b]{0.48\textwidth}
    \centering
    \includegraphics[width=\linewidth]{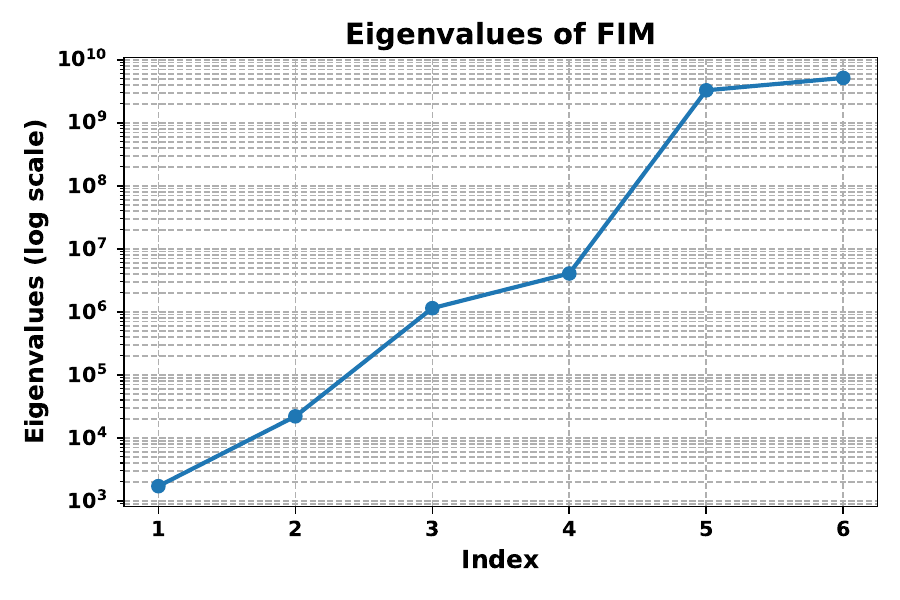}
    \caption{Eigenvalues of FIM}
  \end{subfigure}\hfill
  \begin{subfigure}[b]{0.48\textwidth}
    \centering
    \includegraphics[width=\linewidth]{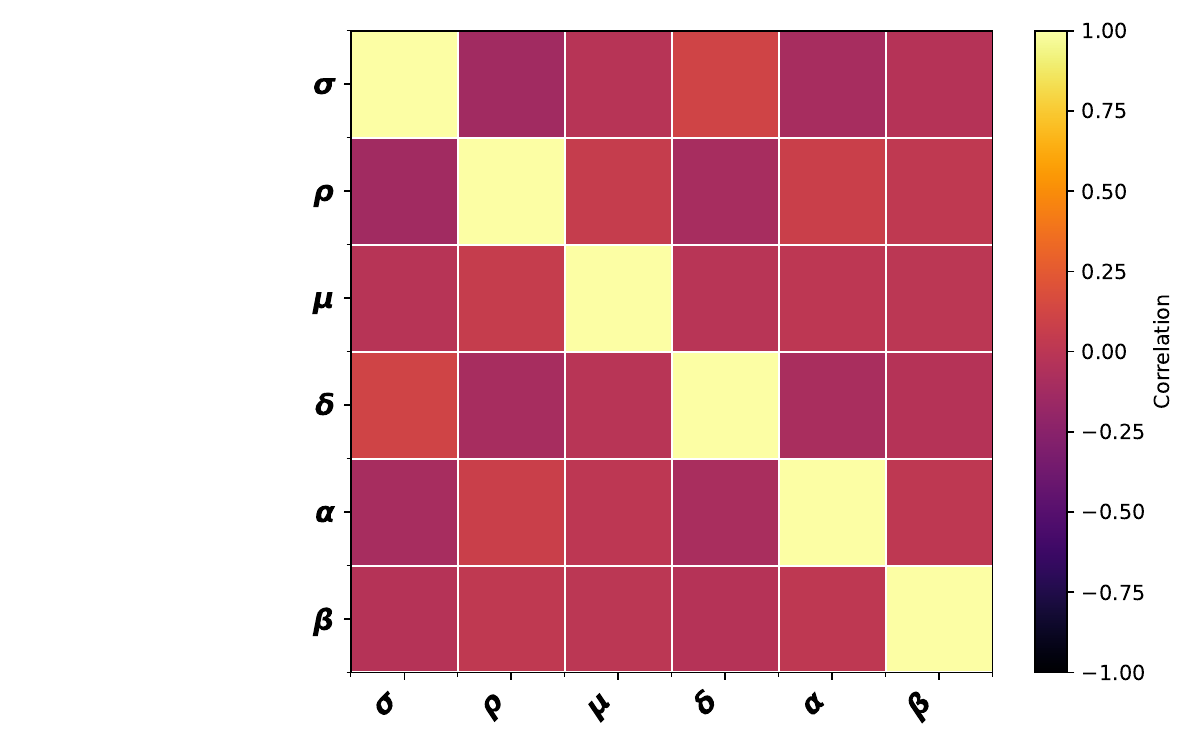}
    \caption{Correlation matrix obtained from FIM}
  \end{subfigure}

  \caption{(Left) Eigenvalues of FIM. (Right) Correlation matrix obtained from FIM of the model (\ref{eq1})-(\ref{eq2}).}
  \label{eigenvalues_and_corelation_model1}
\end{figure}

\vspace{1cm}
We use $5\%$ of the noise level in the measurements to calculate a meaningful FIM. The standard deviation calculated for each parameter for case 1 (only x(t) is observable) is listed in Table \ref{table1_model1}. The eigenvalues and the correlation matrix are shown in Figure \ref{eigenvalues_and_corelation_model1}. 
In this case, no perfect correlations were found between the parameters (except for the diagonal elements). Moreover, the eigenvalues of the FIM are all greater than zero.  This suggests that FIM is invertible and that the standard errors may be computed from its inverse, indicating the local practical identifiability of the parameters. This result is consistent with the parameter inference results given in Table \ref{table1_model1}. Similarly, for case 2, no perfect correlations were found between the parameters indicating local practical identifiability.

For the gray-box identification problem, our objective is to discover the unknown term $f(t)$ of the first ODE of the system (\ref{eq1})-(\ref{eq2}) using sparse data for $x(t)$ and $y(t)$. The hyperparameters of the network are given in Table \ref{table_pinns_hyperparamter}. The performance of the CBINN model is evaluated using the mean absolute error (MAE), root mean square error (RMSE), and relative error (RE) which are defined as follows. 
$$\text{MAE} = \frac{1}{N} \sum_{i=1}^N |\hat{f}(t_i) - f(t_i)|$$

\[
\text{RMSE} = \sqrt{ \frac{ \sum\limits_{i=1}^{N} \left( \hat{f}(t_i) - f(t_i) \right)^2 }{N} },
\]

\[
\text{RE} = \frac{ \sqrt{ \sum\limits_{i=1}^{N} \left( \hat{f}(t_i) - f(t_i) \right)^2 } }
{ \sqrt{ \sum\limits_{i=1}^{N} \hat{f}(t_i)^2 } }
\]

\begin{figure}[H]
\begin{center}
\includegraphics[width=6in, height=2.8in, angle=0]{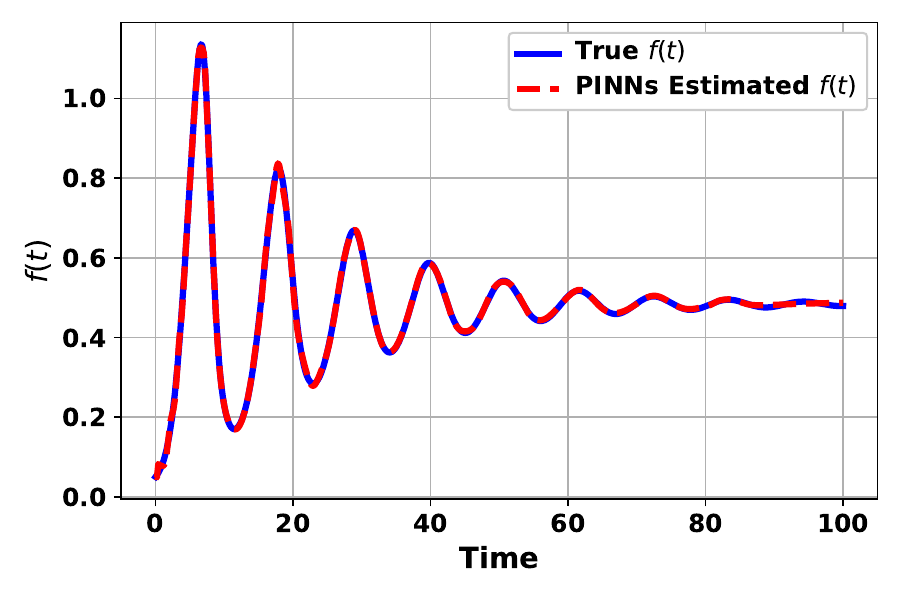}
 \caption{ Plot showing comparison between exact and CBINN estimated $f(t)$ with 50 data points per variable.}
\label{pinns_pred_f}
\end{center}
\end{figure}

\begin{figure}[hbt!]
\begin{center}
\includegraphics[width=8in, height=3.0in, angle=0]{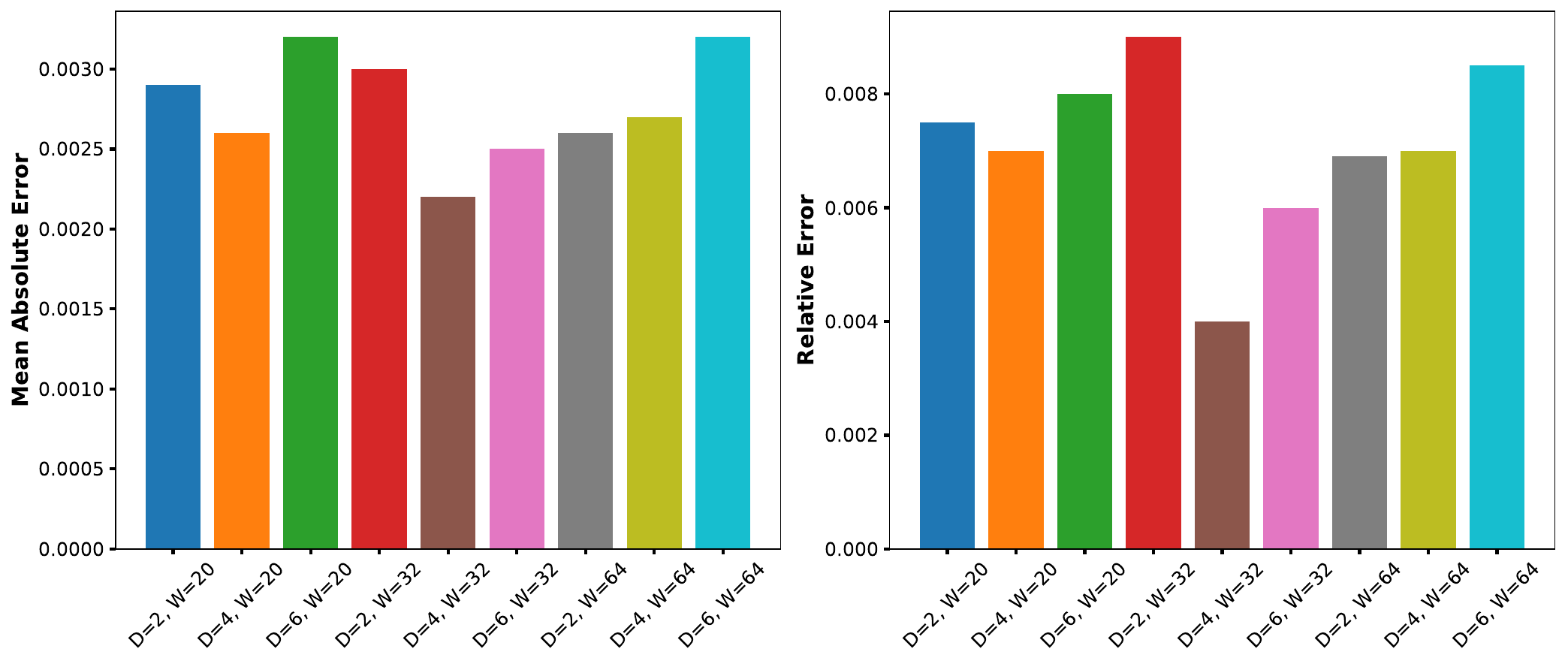}
 \caption{ Mean Absolute Error (left) and Relative Error (right) for discovering $f(t)$ with different network sizes (depth $D$, width $W$) using 100 data points. The CBINN architecture with $D=4,\,W=32$ achieves the best performance across both metrics, while deeper or wider models provide no significant benefit. }
\label{MAE_model1}
\end{center}
\end{figure}

Here, $f(t)$ and $\hat{f}(t)$ are the true and learned solutions, respectively. The expression of true $f(t)$ is $\frac{\rho\,x\,y}{\eta + y}-\mu\,x\,y$. We did not train the model on the full synthetic time series. Instead, we uniformly subsampled the simulation output by taking every $k$-th point (with $k$ equal to the sample rate), yielding a regularly spaced training set that both reduces computational cost and mimics discrete measurement sampling. The CBINN predicted solution and exact solution for unknown $f(t)$  is shown in Figure \ref{pinns_pred_f} using 50 data points for the model variables $x$ and $y$. In Table \ref{tab:f_discovery_combined} (left), we compare the performance of the model with different data sizes. The CBINN model maintains strong performance when trained on small datasets, demonstrating its robustness to data sparsity and its ability to accurately approximate complex dynamics from few observations.  We also evaluate the performance of the models for noisy data, simulating a more realistic scenario. For this we perturb 100 synthetic data points with a uniform random distribution noise at different noise levels. Table \ref{tab:f_discovery_combined} (right) reports MAE, RMSE, and RE($\%$) for discovering the missing term $f(t)$ at different noise levels. The CBINN discovers $f(t)$ accurately in all cases; errors remain small and increase with noise, as expected. In Figure \ref{MAE_model1}, we compared nine CBINN architectures with depth $D\in\{2,4,6\}$ and width $W\in\{20,32,64\}$ using 100 data points. Both Mean Absolute Error (MAE) and Relative Error (RE) are lowest for $D=4,\;W=32$. Increasing depth to $D=6$ or width to $W=64$ is not found to significantly improve the accuracy, a moderate size model is sufficient to discover $f(t)$ in this case.

\begin{table}[h!]
\centering
\captionsetup{justification=centering}
\caption{Unknown term discovery ($f(t)$) over [0, 100] days. CBINN performance (MAE, RMSE, RE). Left: varying number of data points. Right: varying noise with number of data points fixed at 100.}
\renewcommand{\arraystretch}{1.3}

\begin{subtable}[t]{0.48\textwidth}
\centering
\caption{Varying number of data points}
\begin{tabular}{|c||c|c|c|}
\hline \hline
\textbf{No. of Data Points} & \multicolumn{3}{c|}{$\mathbf{f(t)}$} \\
\cline{2-4}
& \textbf{MAE} & \textbf{RMSE} & \textbf{RE ($\%$)} \\
\hline
250  & $1.7 \times 10^{-3}$  & $2.1 \times 10^{-3}$  & 0.1 \\
\hline \hline
100  & $2.3 \times 10^{-3}$  & $3.1 \times 10^{-3}$  & 0.4\\
\hline \hline
50   & $5.7 \times 10^{-3}$  & $8.9 \times 10^{-3}$  & 1.9 \\
\hline \hline
30   & $9.8 \times 10^{-3}$  & $2.4 \times 10^{-2}$  & 6.5\\
\hline \hline
\end{tabular}
\end{subtable}\hfill
\begin{subtable}[t]{0.48\textwidth}
\centering
\caption{Varying noise level ( No. of data points = 100)}
\begin{tabular}{|c||c|c|c|}
\hline \hline
\textbf{Noise Level} & \multicolumn{3}{c|}{$\mathbf{f(t)}$} \\
\cline{2-4}
& \textbf{MAE} & \textbf{RMSE} & \textbf{RE ($\%$)} \\
\hline
1\%   & $2.5 \times 10^{-3}$  & $3.1 \times 10^{-3}$  & 0.45 \\
\hline \hline
2\%   & $2.86 \times 10^{-3}$  & $3.8 \times 10^{-3}$  & 0.75 \\
\hline \hline
5\%   & $7.1 \times 10^{-3}$  & $1.0 \times 10^{-2}$  & 2.2 \\
\hline \hline
10\%  & $1.62 \times 10^{-2}$  & $ 2.84 \times 10^{-2}$  & 5.6 \\
\hline \hline
\end{tabular}
\end{subtable}

\label{tab:f_discovery_combined}
\end{table}

\subsection{\textbf{Tumor-Immune Model with Cytokine}}
\begin{figure}[hbt!]
\begin{center}
\includegraphics[width=10.5in, height=5in, angle=0]{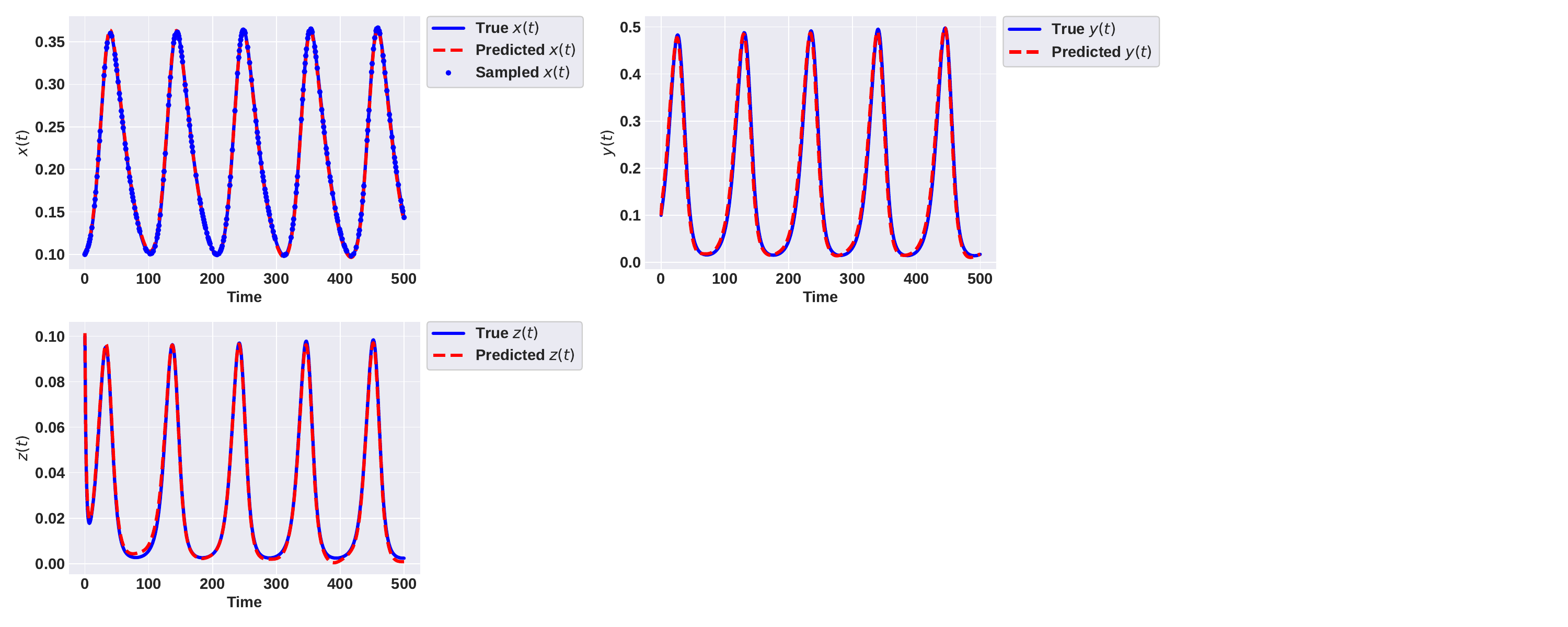}
 \caption{ CBINN inferred dynamics of the model (\ref{eq25})-(\ref{eq28}) compared with the exact solution. 200 noiseless measurements of $x(t)$ were sampled from synthetically generated data and used to infer the model parameters and reconstruct the dynamics of the unobserved states.}
\label{fig1_model2}
\end{center}
\end{figure}

\begin{figure}[H]
\begin{center}
\includegraphics[width=7.5in, height=2.5in, angle=0]{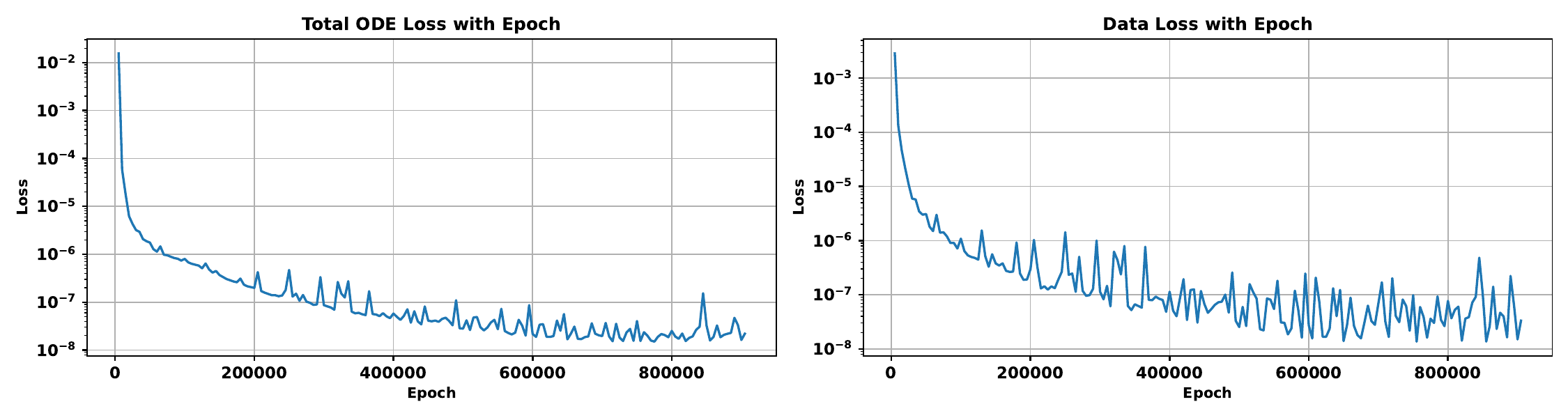}
 \caption{Total ODE residual (sum of the three equation residuals) and data loss (right) plotted against iterations when only $x(t)$ is observable for the model (\ref{eq25})–(\ref{eq28}).}
\label{ode_data_loss_model2}
\end{center}
\end{figure}

\begin{figure}[H]
\begin{center}
\includegraphics[width=7.5in, height=2.5in, angle=0]{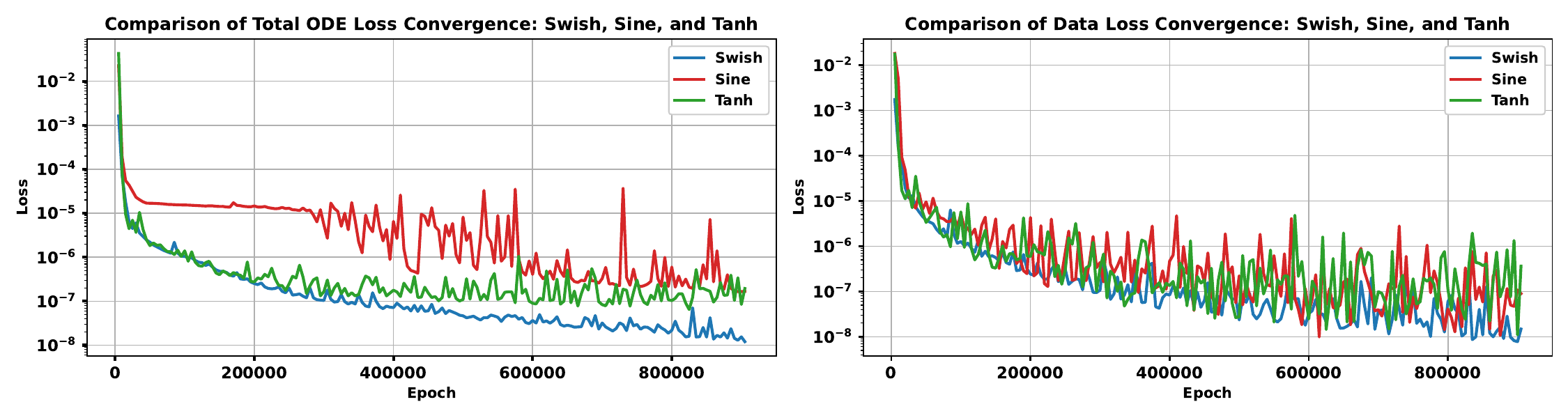}
 \caption{ Loss histories for model (\ref{eq25})-(\ref{eq28}) : total ODE residual (sum over the three equations; left) and data loss (right) versus iterations for Swish, Tanh, and Sine activations. Swish converges fastest and to the lowest losses with minimal variance; tanh attains intermediate performance; sine exhibits larger oscillations and higher residuals. }
\label{loss_compare_model2}
\end{center}
\end{figure}

Tumor-immune model with cytokine introduces qualitatively new dynamics (multiple saturation terms, and an additional state variable) that are absent in the previous model, making it a rigorous test of the CBINN's capacity to learn complex behavior. In practical terms, accurately fitting this system requires estimating more parameters and capturing the interaction between all compartments; using it as a test case therefore directly assesses the robustness and flexibility of the CBINN under nontrivial conditions. Here, we first aim to infer the parameters and the hidden dynamics of the unobserved variables of the model (\ref{eq25})-(\ref{eq28}). The synthetic data are generated by solving the ODE (\ref{eq25})-(\ref{eq28}) using the LSODA algorithm implemented in Python’s scipy.integrate.odeint function. Numerical simulations were carried out over the time interval $t\in[0,500]$ days with the parameter set
\[
c=0.035,\ \mu_2=0.03,\ p_1=0.1245,\ r_2=0.18,\ b=0.002,\ a=1,\ p_2=0.5,\ \mu_3=0.5,\ g_1=g_2=g_3=1,
\]
and initial condition $(x,y,z)=(0.1,0.1,0.1)$. The numerical solutions generated using LSODA algorithm is shown in Figure \ref{fig1_model2}. The model exhibit oscillatory dynamics with this chosen set of parameter values. We consider two scenarios in this model. In the first, we consider $x(t)$ as the only observable variable, fix the parameters $g_1, g_2$ and $g_3$ to their target values and infer rest of the parameters. In the second case, we infer all the parameters assuming all model variables are observable. To test robustness, measurements were corrupted with zero-mean Gaussian noise at 2\%, 5\% and 10\% levels. Using the CBINN model and the input data, we are able to accurately discover the parameters and also correctly infer the dynamics of the unobserved variables of the model (\ref{eq25})-(\ref{eq28}) as shown in Figure \ref{fig1_model2}. The true and estimated parameter values along with the mean absolute errors are shown in Table \ref{table1_mode2}. These results show a good agreement between the inferred and exact dynamics using just one observable variable. The total ode loss (sum of three odes) and the data loss for this case with noiseless observations are shown in Figure \ref{ode_data_loss_model2}. The comparison between three different activation functions is shown in Figure \ref{loss_compare_model2}. We find that swish converges fastest and to the lowest losses with minimal variance.  \\

\begin{figure}[hbt!]
  \centering
  \begin{subfigure}[b]{0.48\textwidth}
    \centering
    \includegraphics[width=\linewidth]{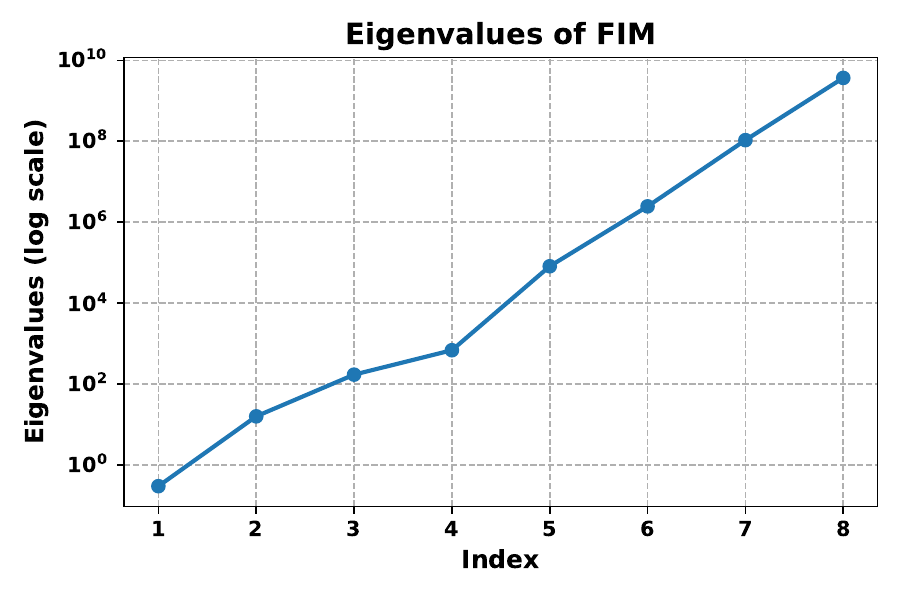}
    \caption{Eigenvalues of FIM}
  \end{subfigure}\hfill
  \begin{subfigure}[b]{0.48\textwidth}
    \centering
    \includegraphics[width=\linewidth]{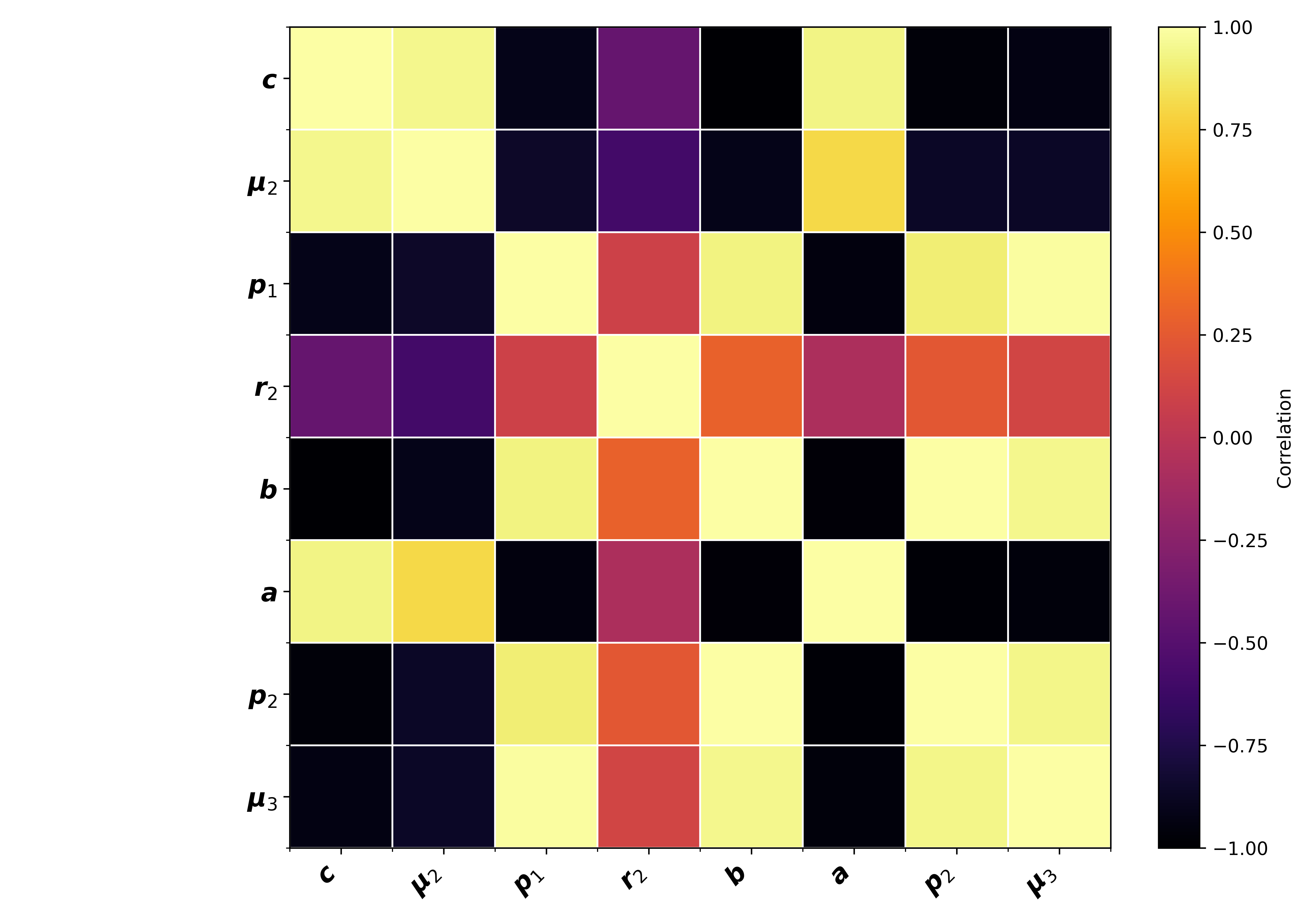}
    \caption{Correlation matrix obtained from FIM}
  \end{subfigure}

  \caption{(Left) Eigenvalues of FIM. (Right) Correlation matrix obtained from FIM of model (\ref{eq25}) - (\ref{eq28}).}
  \label{eigenvalues_and_corelation_model2}
\end{figure}

\begin{figure}[H]
\begin{center}
\includegraphics[width=10.5in, height=3.5in, angle=0]{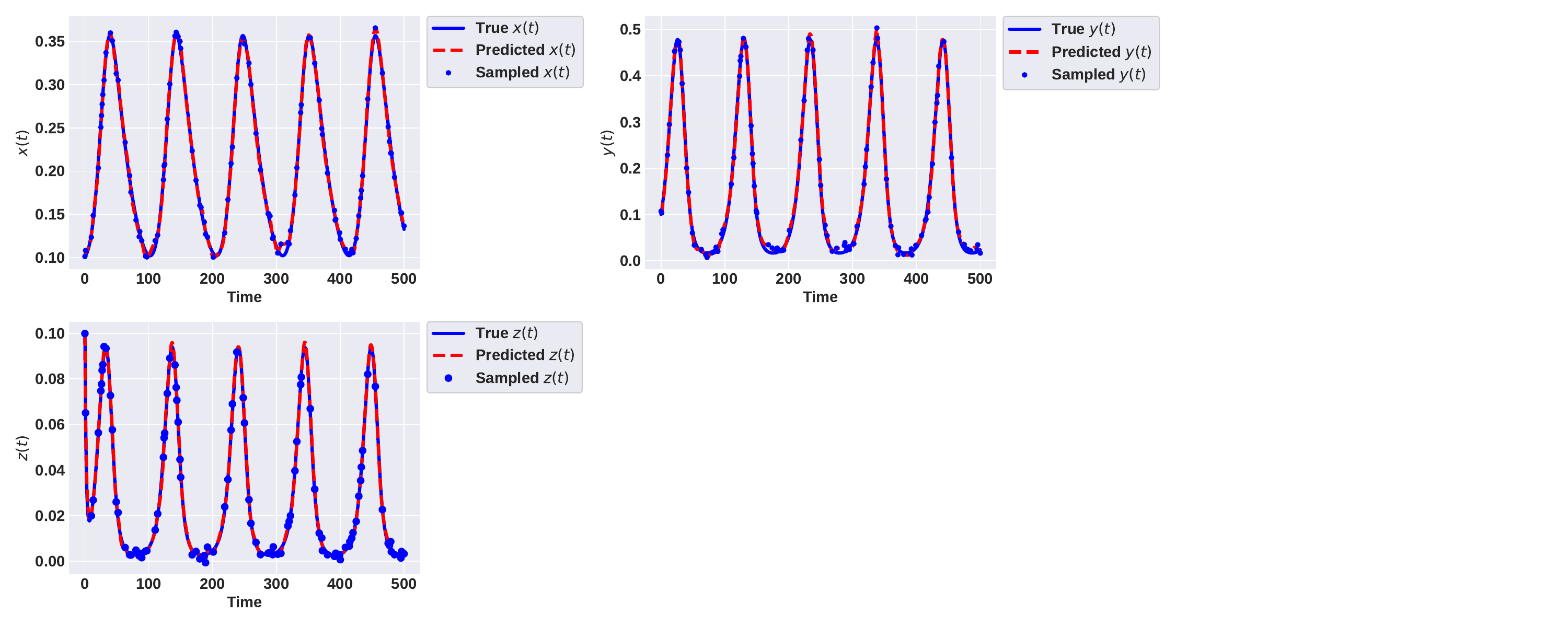}
 \caption{ Inferred dynamics of tumor-immune model with cytokine (\ref{eq25})-(\ref{eq28}) from  noisy observations compared with the exact solution. Predictions are
 performed on equally-spaced time instants in the interval of $0 - 500$ days. 100 measurements from each x, y and z are corrupted by a zero-mean Gaussian noise and standard deviation of $\sigma^* = 0.05c$ where $c$ is the standard deviation of each observable variable over the observation time window. }
\label{fig2model2}
\end{center}
\end{figure}

In Figure \ref{eigenvalues_and_corelation_model2}, the eigenvalues and correlation matrix computed from FIM using $5\%$ noise on $x$ measurements is shown. We find no perfect correlations between the parameters, except for $\beta$ and $p_2$. The corelation value between $\beta$ and $p_2$ were found to be $0.989$. This suggests that with higher noise and $x(t)$ as the only observable variable $\beta$ and $p_2$ may not be practically identifiable. These findings are also consistent with the estimated values and the standard deviation reported in Table \ref{table1_mode2} (standard deviation greater than the estimated values and large error with increasing noise levels). To complement the correlation analysis, we examined the eigenvalue spectrum of the Fisher information matrix (FIM). The smallest eigenvalue is $\lambda_{\min}=0.163$ and the corresponding eigenvector is dominated by the $p_2$ component, indicating that $p_2$ (or a parameter combination involving $p_2$) is weakly informed by the data and is practically non-identifiable. The CBINN inferred dynamics for the second case is presented in Figure \ref{fig2model2}. In this case, we took 100 measurements from each state variable and used CBINN to infer all the parameters. The inferred parameter values closely align with the true values ( Table \ref{table2_mode2}). The FIM analysis showed no perfect correlations among the parameters, implying their practical identifiability.\\

\begin{table}[tbp]
  \centering
  \caption{Case 1: only $x$ is observable, $g_i$ for $1\leq i\leq 3$ are fixed. Target values, CBINN estimates, mean absolute errors, and (empty) standard-deviation column for the tumor–immune cytokine model (\ref{eq25})-(\ref{eq28}) under noiseless and noisy measurements. The final column is left blank for the standard deviations (to be filled later).}
  \label{table1_mode2}
  \setlength{\tabcolsep}{4pt}
  \small
  \begin{tabular}{l
                  S[table-format=1.5]     
                  *{4}{S[table-format=1.5]}
                  *{4}{S[table-format=1.5]}
                  S[table-format=1.5]     
                  }
    \toprule
    & & \multicolumn{4}{c}{\textbf{Estimated value}} & \multicolumn{4}{c}{\textbf{Mean Absolute error}} & \multicolumn{1}{c}{\textbf{Std.\ dev.}} \\
    \cmidrule(lr){3-6}\cmidrule(lr){7-10}\cmidrule(lr){11-11}
    \textbf{Parameter} & \textbf{Target}
      & \textbf{Noiseless} & \textbf{2\%} & \textbf{5\%} & \textbf{10\%}
      & \textbf{Noiseless} & \textbf{2\%} & \textbf{5\%} & \textbf{10\%}
      & \textbf{} \\
    \midrule
    $c$      & 0.0350 & 0.0323 & 0.0345 & 0.045 & 0.038 & 0.0028 & 0.00045 & 0.01 & 0.0032 & {0.0016} \\
    $\mu_2$  & 0.0300 & 0.0302 & 0.0305 & 0.0306 & 0.036 & 0.00021 & 0.00052 & 0.00065 & 0.005 & {0.0003} \\
    $p_1$    & 0.1245 & 0.1221 & 0.1201 & 0.1191 & 0.132 & 0.0023 & 0.0045 & 0.0055 & 0.0076  & {0.0755} \\
    $r_2$    & 0.1800 & 0.1770 & 0.1760 & 0.1790 & 0.156& 0.0021 & 0.008 & 0.001 & 0.0256 & {0.0015} \\
    $\beta$  & 0.0020 & 0.0032 & 0.0098 & 0.0120 & 0.032 & 0.0015 & 0.0074 & 0.015 & 0.032&{0.1650} \\
    $a$      & 1.0000 & 0.9990 & 0.9820 & 0.9800 & 0.896 & 0.001 & 0.0180 & 0.020 & 0.050 & {0.0441} \\
    $p_2$    & 0.5000 & 0.4830 & 0.4720 & 0.251 & 0.125 & 0.018 & 0.0301 & 0.192 &  0.352 & {0.5271} \\
    $\mu_3$  & 0.5000 & 0.4740 & 0.4680 & 0.3321 & 0.285 & 0.026 & 0.033 & 0.165 & 0.220 & {0.2120} \\
    \bottomrule
  \end{tabular}
\end{table}

\begin{figure}[H]
\begin{center}
\includegraphics[width=8in, height=2.5in, angle=0]{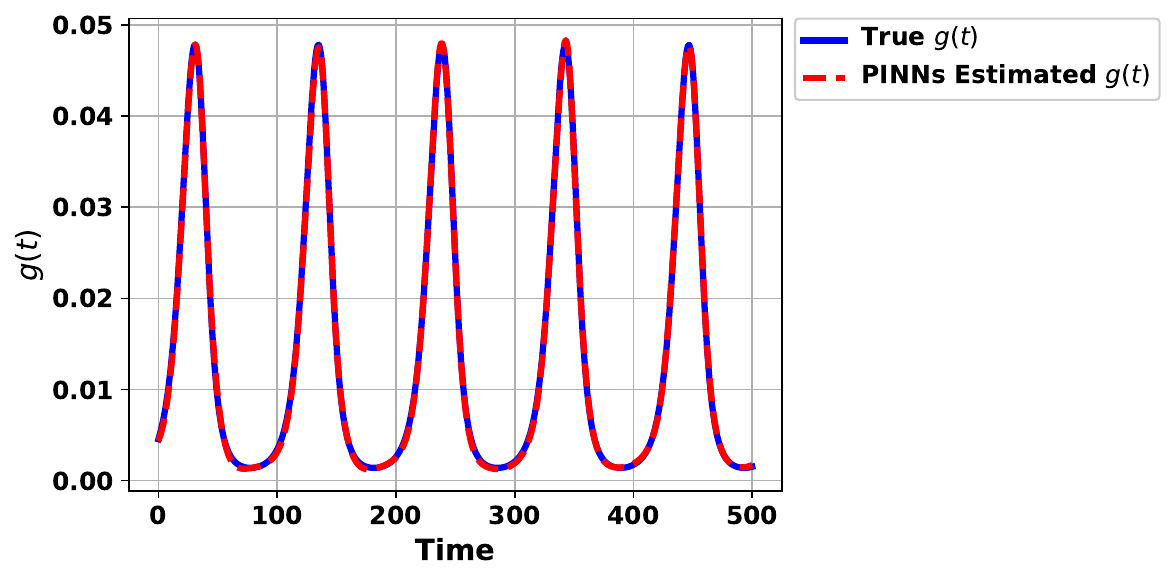}
 \caption{ Plot showing comparison between exact and PINNs solutions for the unknown term $g(t)$ with 100 data points per variable.}
\label{pinns_pred_g}
\end{center}
\end{figure}

\begin{figure}[H]
\begin{center}
\includegraphics[width=8in, height=3.2in, angle=0]{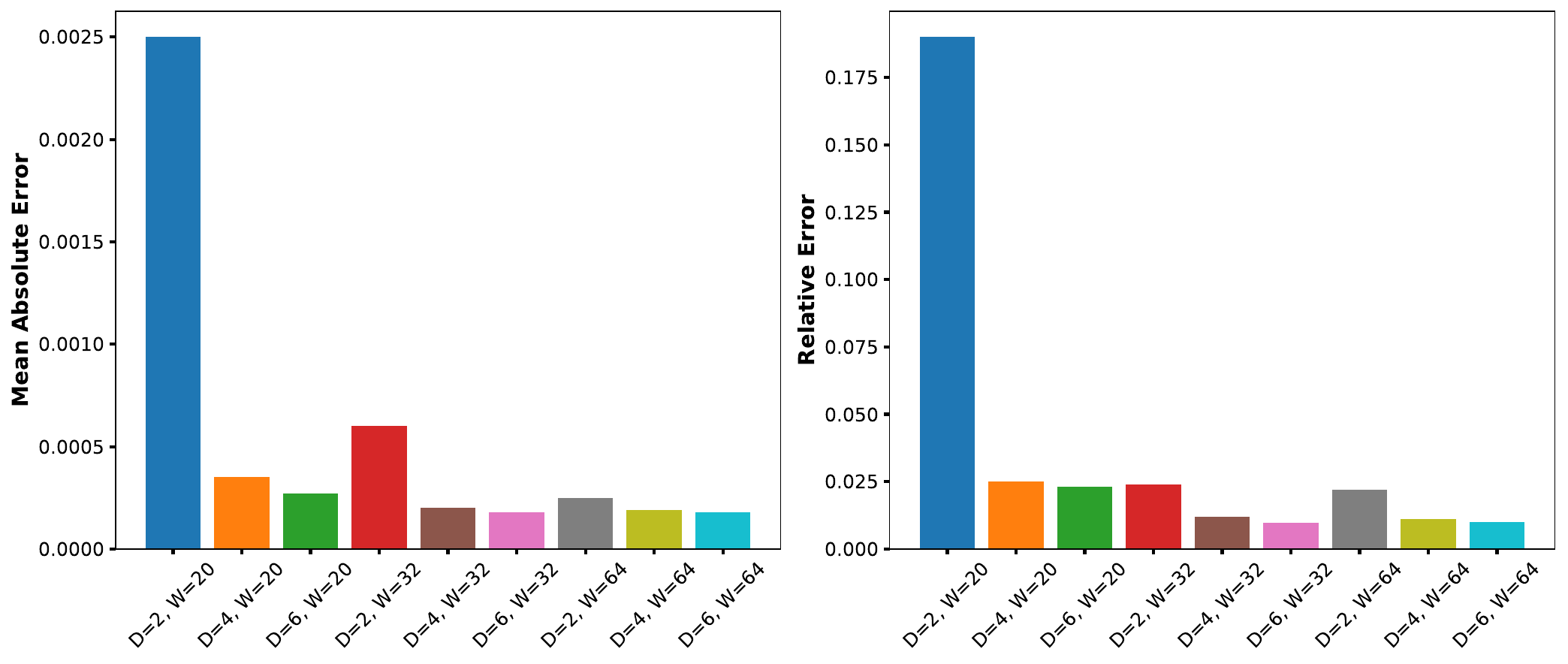}
 \caption{ Mean Absolute Error (left) and Relative Error (right) for recovering $g(t)$ using 100 data points across nine CBINN architectures (depth $D$, width $W$).
    Errors are lowest for $D = 4-6$ with $W = 32$, while shallow models ($D=2$) especially $D=2,\,W=20$ exhibit substantially higher error. }
\label{MAE_model2}
\end{center}
\end{figure}

\vspace{1cm}
For the gray-box identification problem, we aim to discover the missing term $g(t)$ using the measurement data of the model variables $x$, $y$ and $z$. The performance of CBINN model is evaluated using mean absolute error (MAE), root mean square error (RMSE), and relative error (RE). The residual loss is evaluated over 10000 points (collocation points). The CBINN predicted solution and the exact numerical solution for unknown $g(t)$  is presented in Figure \ref{pinns_pred_g}. From the figure, we clearly see that the predicted solutions exactly overlaps with the exact one. The comparison of the CBINN performance with varying data sizes and different noise levels is presented in Table \ref{tab:g_discovery_combined}. As shown in Table~\ref{tab:g_discovery_combined}, increasing the sample size from 30 to 250 points reduces the error monotonically in discovering $g(t)$. Model performance improves with more data and gradually degrades with higher data noise. In Figure \ref{MAE_model2}, MAE and RE are compared with varying depth and width of the CBINN architecture using 100 data points. Both MAE and RE drop when moving from a shallow ($D{=}2$) to moderate depth ($D{=}4$) architecture. The lowest errors are achieved for $D\in\{4,6\}$ with $W = 32$. This indicates that a moderate depth/width provides the most reliable recovery of $g(t)$.

\begin{table}[tbp]
  \centering
  \caption{Case 2: All three variables $(x, y, z)$ are observable: target values, CBINN estimates, and absolute errors for the tumor–immune cytokine model (\ref{eq25})-(\ref{eq28}) under noiseless and noisy measurements.}
  \label{table2_mode2}
  \setlength{\tabcolsep}{4pt}
  \small
  \begin{tabular}{l
                  S[table-format=1.4]
                  *{4}{S[table-format=1.4]}
                  *{4}{S[table-format=1.4]}}
    \toprule
    & & \multicolumn{4}{c}{\textbf{Estimated value}} & \multicolumn{4}{c}{\textbf{Mean Absolute error}} \\
    \cmidrule(lr){3-6}\cmidrule(lr){7-10}
    \textbf{Parameter} & \textbf{Target}
      & \textbf{Noiseless} & \textbf{2\%} & \textbf{5\%} & \textbf{10\%}
      & \textbf{Noiseless} & \textbf{2\%} & \textbf{5\%} & \textbf{10\%} \\
    \midrule
    $c$      & 0.0350 & 0.0351 & 0.0353 & 0.0355 & 0.0356 & 0.0001 & 0.0003 & 0.0005 & 0.0006 \\
    $\mu_2$  & 0.0300 & 0.0303 & 0.0302 & 0.0306 & 0.0303 & 0.0003 & 0.0002 & 0.0006 & 0.0003 \\
    $p_1$    & 0.1245 & 0.1250 & 0.1120 & 0.1140 & 0.2120 & 0.0005 & 0.0125 & 0.0105 & 0.0875 \\
    $r_2$    & 0.1800 & 0.1790 & 0.1780 & 0.1790 & 0.1820 & 0.0010 & 0.0020 & 0.0010 & 0.0020 \\
    $\beta$  & 0.0100 & 0.0120 & 0.0098 & 0.0170 & 0.0210 & 0.0020 & 0.0002 & 0.0070 & 0.0110 \\
    $a$      & 1.0000 & 0.9980 & 0.9880 & 0.9780 & 0.9650 & 0.0020 & 0.0120 & 0.0220 & 0.0350 \\
    $p_2$    & 0.5000 & 0.5040 & 0.5090 & 0.4830 & 0.4810 & 0.0040 & 0.0090 & 0.0170 & 0.0190 \\
    $\mu_3$  & 0.5000 & 0.4870 & 0.4860 & 0.4670 & 0.4210 & 0.0130 & 0.0140 & 0.0330 & 0.0790 \\
    $g_1$    & 1.0000 & 1.0000 & 0.9820 & 0.9420 & 1.2130 & 0.0000 & 0.0180 & 0.0580 & 0.2130 \\
    $g_2$    & 1.0000 & 0.9870 & 0.9820 & 0.9720 & 0.9310 & 0.0130 & 0.0180 & 0.0280 & 0.0690 \\
    $g_3$    & 1.0000 & 1.0051 & 1.0620 & 1.0540 & 1.2100 & 0.0051 & 0.0620 & 0.0540 & 0.2100 \\
    \bottomrule
  \end{tabular}
\end{table}

\begin{table}[h!]
\centering
\captionsetup{justification=centering}
\caption{Unknown term discovery ($g(t)$) over $t\in[0,500]$ days. CBINN performance (MAE, RMSE, RE). Left: effect of sample size. Right: effect of noise with number of data points fixed at $N=100$.}
\renewcommand{\arraystretch}{1.3}

\begin{subtable}[t]{0.48\textwidth}
\centering
\caption{Varying number of data points}
\label{subtab:g_points}
\begin{tabular}{|c||c|c|c|}
\hline \hline
\textbf{No.~of data points} & \multicolumn{3}{c|}{$\mathbf{g(t)}$} \\
\cline{2-4}
& \textbf{MAE} & \textbf{RMSE} & \textbf{RE ($\%$)} \\
\hline
250  & $1.4.0 \times 10^{-4}$ & $1.9 \times 10^{-4}$ & 0.9 \\
\hline \hline
100  & $2 \times 10^{-4}$ & $2.5 \times 10^{-4}$  & 1.2 \\
\hline \hline
50   & $2.56 \times 10^{-4}$ & $3.2 \times 10^{-4}$  & 1.56 \\
\hline \hline
30   & $7.9 \times 10^{-4}$ & $1.3 \times 10^{-3}$  & 6.54  \\
\hline \hline
\end{tabular}
\end{subtable}\hfill
\begin{subtable}[t]{0.48\textwidth}
\centering
\caption{Varying noise level ($N=100$ data points)}
\label{subtab:g_noise}
\begin{tabular}{|c||c|c|c|}
\hline \hline
\textbf{Noise level} & \multicolumn{3}{c|}{$\mathbf{g(t)}$} \\
\cline{2-4}
& \textbf{MAE} & \textbf{RMSE} & \textbf{RE ($\%$)} \\
\hline
1\%   & $2.25 \times 10^{-4}$  & $3.1 \times 10^{-4}$   & 1.4 \\
\hline \hline
2\%   & $4.3\times 10^{-4}$  & $7.2 \times 10^{-4}$   & 2.5 \\
\hline \hline
5\%   & $8.2 \times 10^{-4}$ & $1.2\times 10^{-3}$  & 5.1 \\
\hline \hline
10\%  & $2.2 \times 10^{-3}$ & $6.7 \times 10^{-3}$ & 8.9 \\
\hline \hline
\end{tabular}
\end{subtable}

\label{tab:g_discovery_combined}
\end{table}

\subsection{\textbf{Tumor - Immune Model with Immune Cell Conversion}}

\begin{figure}[hbt!]
\begin{center}
\includegraphics[width=7.5in, height=3.2in, angle=0]{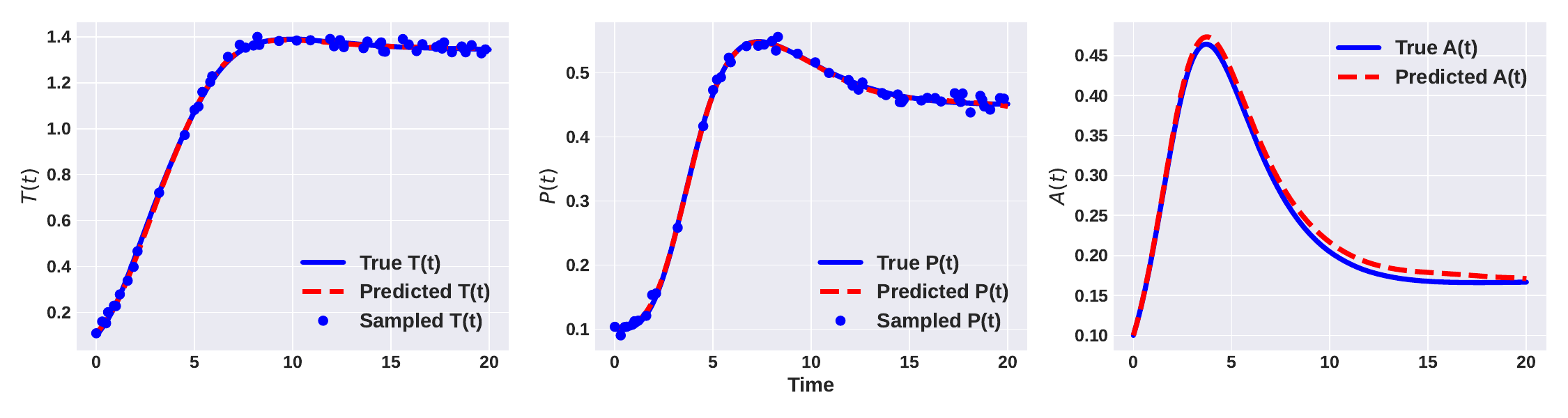}
 \caption{CBINN inferred dynamics of tumor-immune model with cell conversion  (\ref{eq:T})-(\ref{eq:A}) from noisy measurements compared with the exact solution. 50 measurements are corrupted by a zero-mean Gaussian noise and standard deviation of $\sigma = 0.05c$.}
\label{fig1_model3}
\end{center}
\end{figure}

\begin{figure}[hbt!]
\begin{center}
\includegraphics[width=7.5in, height=3.2in, angle=0]{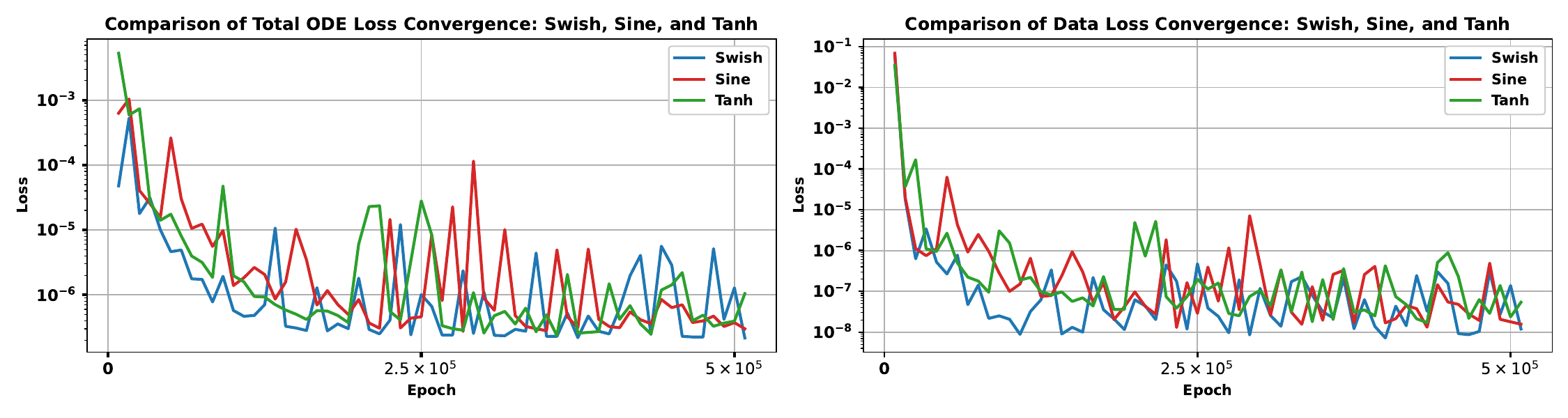}
 \caption{Training curves of total ODE loss (left) and data loss (right) for Swish (SiLU), Sine, and Tanh activations. All activations reach low errors ($\approx 10^{-7}$ ODE, $\approx 10^{-8}$ data), with Swish achieving the most stable and lowest convergence for the model (\ref{eq:T})–(\ref{eq:A}).}
  \label{loss_compare_model3} 
\end{center}
\end{figure}

The synthetic data was generated by simulating the system of equations (\ref{eq:T})-(\ref{eq:A}) with $r_T = 1$, $K_T = 1$, $d_P = 1$, $r_A = 1$, $K_A = 1$, $\alpha_{TP} = 0.95$,  $\alpha_{TA} = 0.5$,  $\alpha_{PT} = 0.15$,  $\alpha_{AT} = 0.05$,  $\alpha_{AP} = 0.5$,  $\omega = 0.5$  and initial condition $(0.1, 0.1, 0.1)$ over the time interval t = 0 to t = 20 (days), with a total of 200 equally spaced time points. 50 noiseless measurements are sampled at random from the generated synthetic data for model variables $T$ and $P$ and used for data loss.  For residual error, 5000 points (collocation points) were sampled at random within the time frame of $0-20$. The CBINN estimated parameter values with different noise levels in the measurement data and the mean absolute error between the predicted values and the target values is presented in Table \ref{table1_model3}. In this model, we fixed parameter $K_A$ to its true value due to structural non-identifiability and estimated the remaining parameters. The true LSODA generated solution, CBINN inferred solutions and the data points used are shown in Figure \ref{fig1_model3}. Training convergence of ODE and data losses using three activation functions are shown in Figure \ref{loss_compare_model3}.  CBINN with the swish activation (blue) attains lower and smoother loss trajectories for both the ODE and data terms compared to sine (red) and tanh (green), indicating better fit to the measurements.\\

\begin{figure}[H]
  \centering
  \begin{subfigure}[b]{0.45\textwidth}
    \centering
    \includegraphics[width=\linewidth]{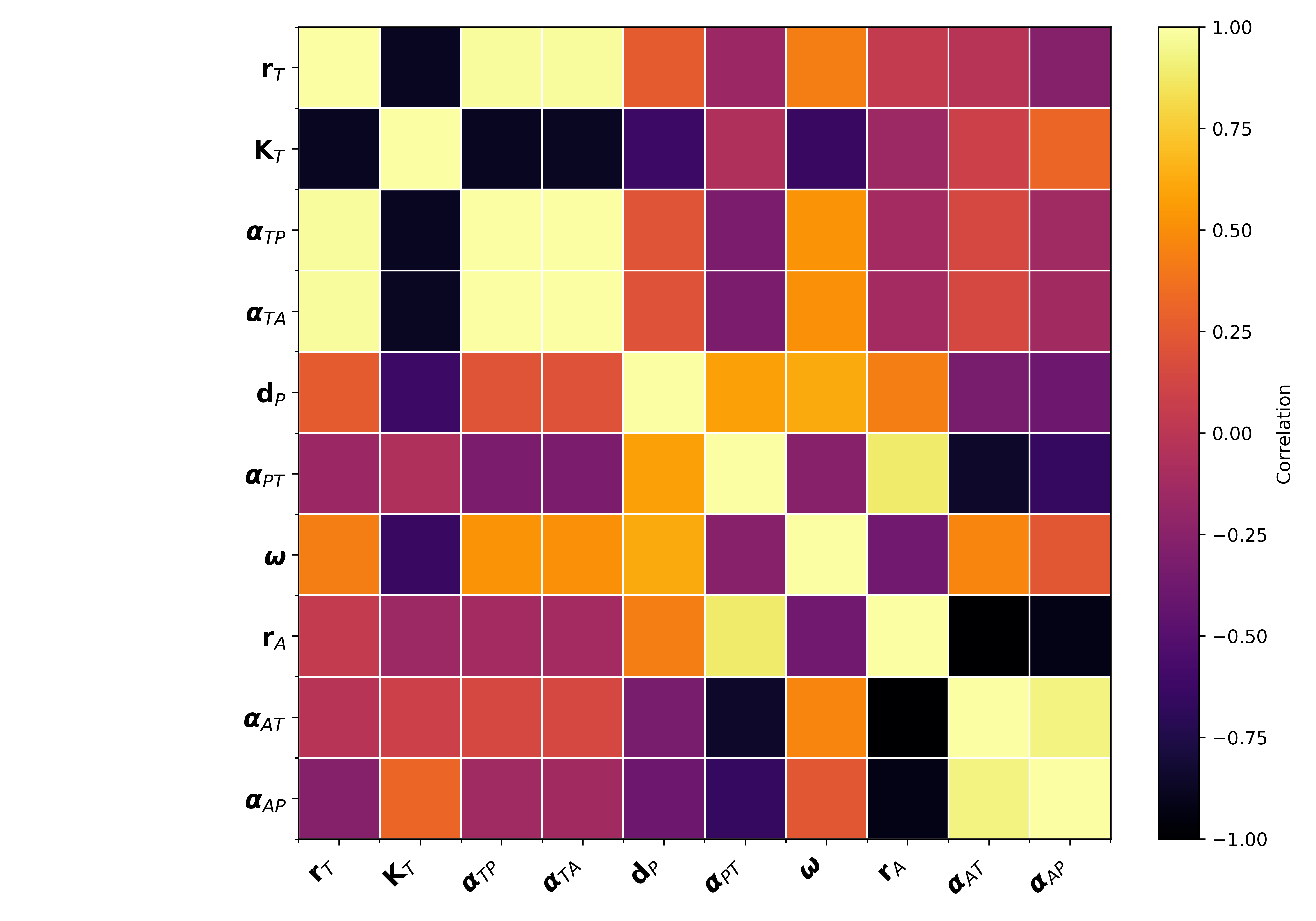}
    \caption{Correlation matrix obtained from FIM}
  \end{subfigure}\hfill
  \begin{subfigure}[b]{0.45\textwidth}
    \centering
    \includegraphics[width=\linewidth]{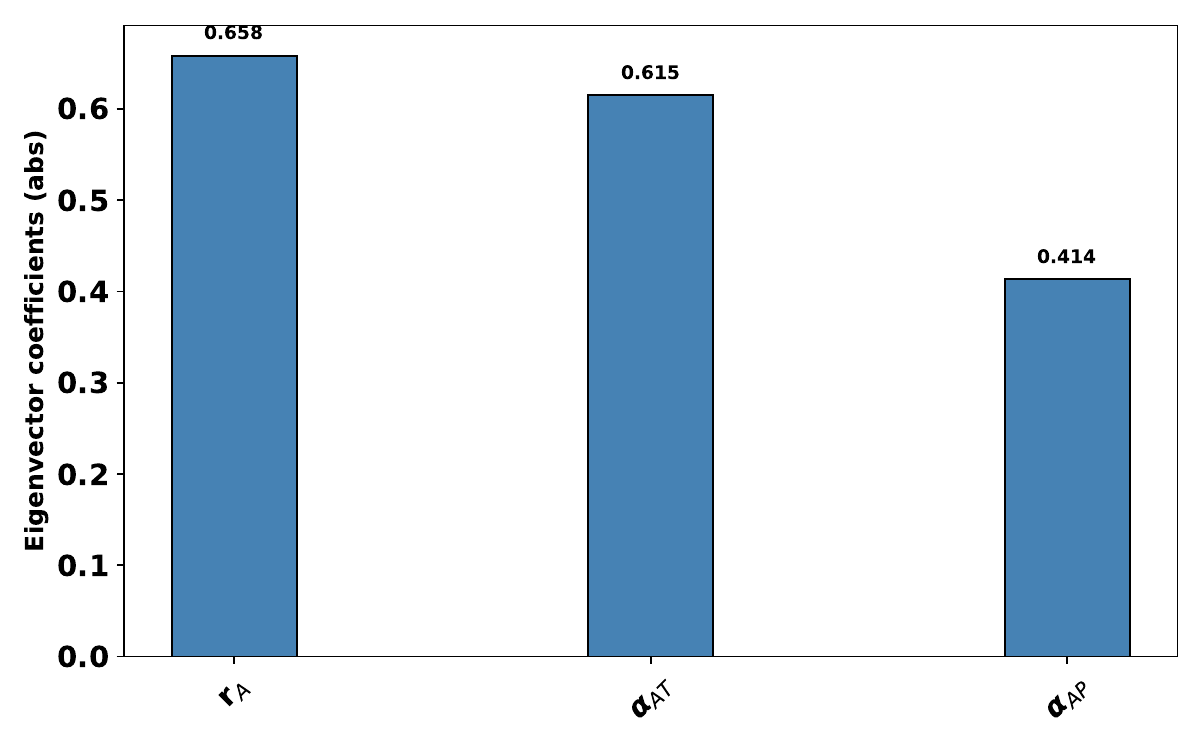}
    \caption{Dominant components of the eigenvector associated with the smallest eigenvalue of the FIM}
  \end{subfigure}

  \caption{(Left) Corelation Matrix. (Right) Dominant components of the eigenvector associated with the smallest eigenvalue of the FIM of model (\ref{eq25}) - (\ref{eq28}).}
  \label{corelation_matrix_model3}
\end{figure}

To test the practical identifiability of the parameters, we have plotted the correlation matrix in Figure \ref{corelation_matrix_model3} (left). The noise level of $5 \%$ was added to the measurements of the observable variables and corelation matrix was computed using FIM. We observe perfect correlations (corelation above 0.99) between parameters $\alpha_{TP} - \alpha_{TA}$ and $r_{A} - \alpha_{AT}$. This indicates that these parameters may not be practically identifiable. To further investigate, we computed the eigenvalues of the FIM. The smallest eigenvalue was found to be 0.69, and we analyzed the corresponding eigenvector. Figure \ref{corelation_matrix_model3} (right) shows the dominant components of this eigenvector. We find that $r_{A}$, $\alpha_{AT}$, and $\alpha_{AP}$ are dominant, while the contributions of other parameters are close to zero. These parameters can therefore be considered practically non-identifiable. Our CBINN estimation results (\ref{table1_model3}) show that the parameter $\alpha_{AT}$ may not be estimated accurately especially when the noise level in the measurements increases, which is consistent with the findings from the FIM analysis. In contrast, our algorithm is able to infer the parameters $r_{A}$ and $\alpha_{AP}$ with good accuracy, even under high noise levels. This result does not align with the conclusions drawn from the FIM analysis. A possible reason is that assessing practical identifiability using the FIM can be challenging, particularly in partially observed nonlinear systems. These observations are consistent with previous findings reported in \cite{yazdani2020systems, joshi2006exploiting}. Alternative approaches, such as bootstrapping methods \cite{balsa2008computational} or probabilistic frameworks \cite{joshi2006exploiting, foo2009multi}, could be used in such scenarios to obtain more reliable information about practical identifiability.

\begin{table}[hbt!]
  \centering
  \caption{Targets, CBINN estimates, and absolute errors for the ODE model (\ref{eq:T})–(\ref{eq:A}) under noiseless and noisy measurements.}
  \label{table1_model3}
  \setlength{\tabcolsep}{4pt}
  \small
  \begin{tabular}{l
                  S[table-format=1.3]
                  *{4}{S[table-format=1.3]}
                  *{4}{S[table-format=1.3]}
                  c} 
    \toprule
    & & \multicolumn{4}{c}{\textbf{Estimated value}} & \multicolumn{4}{c}{\textbf{Mean Absolute error}} & \multicolumn{1}{c}{\textbf{Std.\ dev.}} \\
    \cmidrule(lr){3-6}\cmidrule(lr){7-10}\cmidrule(lr){11-11}
    \textbf{Parameter} & \textbf{Target}
      & \textbf{Noiseless} & \textbf{2\%} & \textbf{5\%} & \textbf{10\%}
      & \textbf{Noiseless} & \textbf{2\%} & \textbf{5\%} & \textbf{10\%}
      & \textbf{} \\
    \midrule
    $r_T$        & 1.000 & 1.000 & 0.980 & 0.890 & 0.856 & 0.0001 & 0.015 & 0.121 & 0.132 & 0.246\\
    $K_T$        & 1.000 & 0.996 & 1.010 & 1.020 & 1.080 & 0.004 & 0.010 & 0.020 & 0.080  & 0.0248\\
    $d_P$        & 1.000 & 1.000 & 1.000 & 1.100 & 1.170 & 0.0001 & 0.0001 & 0.100 & 0.170  & 0.115\\
    $r_A$        & 1.000 & 1.000 & 0.995 & 0.998 & 1.010 & 0.0001 & 0.006 & 0.002 & 0.010 & 0.635\\
    $K_A$        & 1.000 & \text{-} &  \text{-} & \text{-} & \text{-} &  \text{-} & \text{-} & \text{-} & \text{-} & \text{-} \\
    $\alpha_{TP}$& 0.950 & 0.946 & 0.862 & 0.821 & 0.568 & 0.0045 & 0.089 & 0.129 & 0.382 & 0.421\\
    $\alpha_{TA}$& 0.500 & 0.497 & 0.432 & 0.366 & 0.254 & 0.0032 & 0.068 & 0.134 & 0.246 & 0.408\\
    $\alpha_{PT}$& 0.150 & 0.145 & 0.147 & 0.138 & 0.135 & 0.005 & 0.003 & 0.012 & 0.015 & 0.0913\\
    $\alpha_{AT}$& 0.050 & 0.061 & 0.095 & 0.005 & 0.0009 & 0.011 & 0.046 & 0.045 & 0.049 & 0.593\\
    $\alpha_{AP}$& 0.500 & 0.520 & 0.478 & 0.398 & 0.321 & 0.020 & 0.021 & 0.102 & 0.179 & 0.425\\
    $\omega$     & 0.500 & 0.499 & 0.502 & 0.512 & 0.520 & 0.001 & 0.002 & 0.013 & 0.021  & 0.083\\
    \bottomrule
  \end{tabular}
\end{table}

\begin{figure}[H]
\begin{center}
\includegraphics[width=8in, height=3.5in, angle=0]{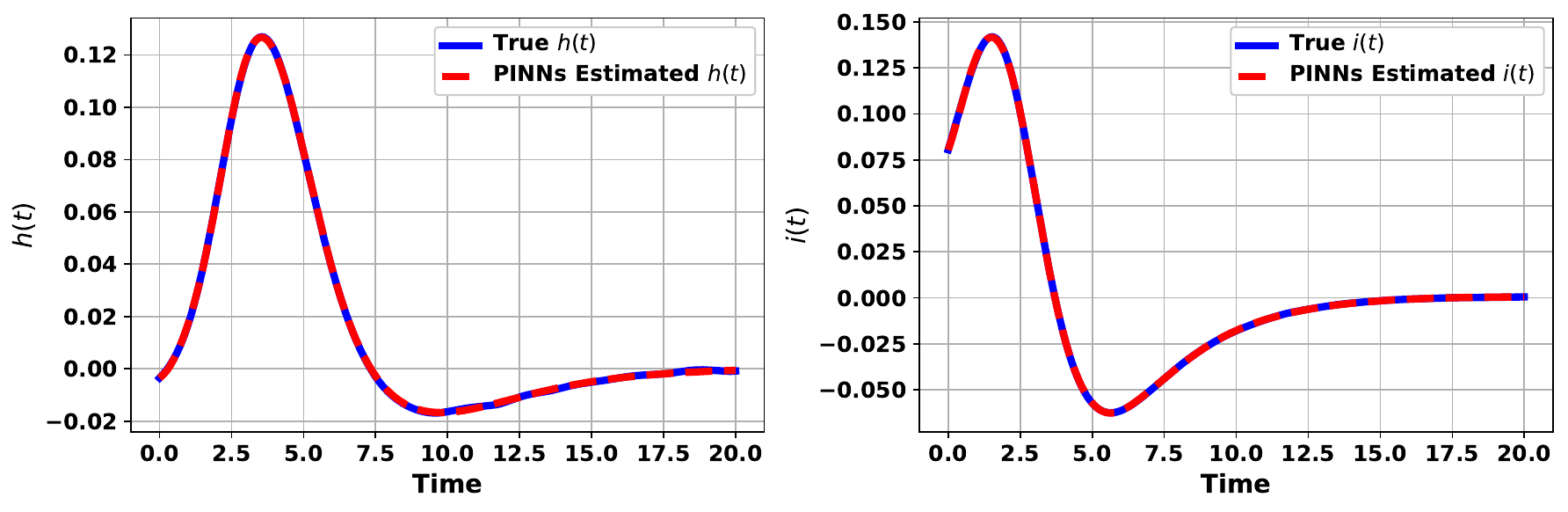}
 \caption{ Plot showing comparison between exact and CBINN estimated solutions for $h(t)$ and $i(t)$ of the model (\ref{eq:T1})–(\ref{eq:A1}). In this case, 30 data points were sampled for each model variable.}
\label{fig2_model3}
\end{center}
\end{figure}

\begin{figure}[H]
\begin{center}
\includegraphics[width=8in, height=3.0in, angle=0]{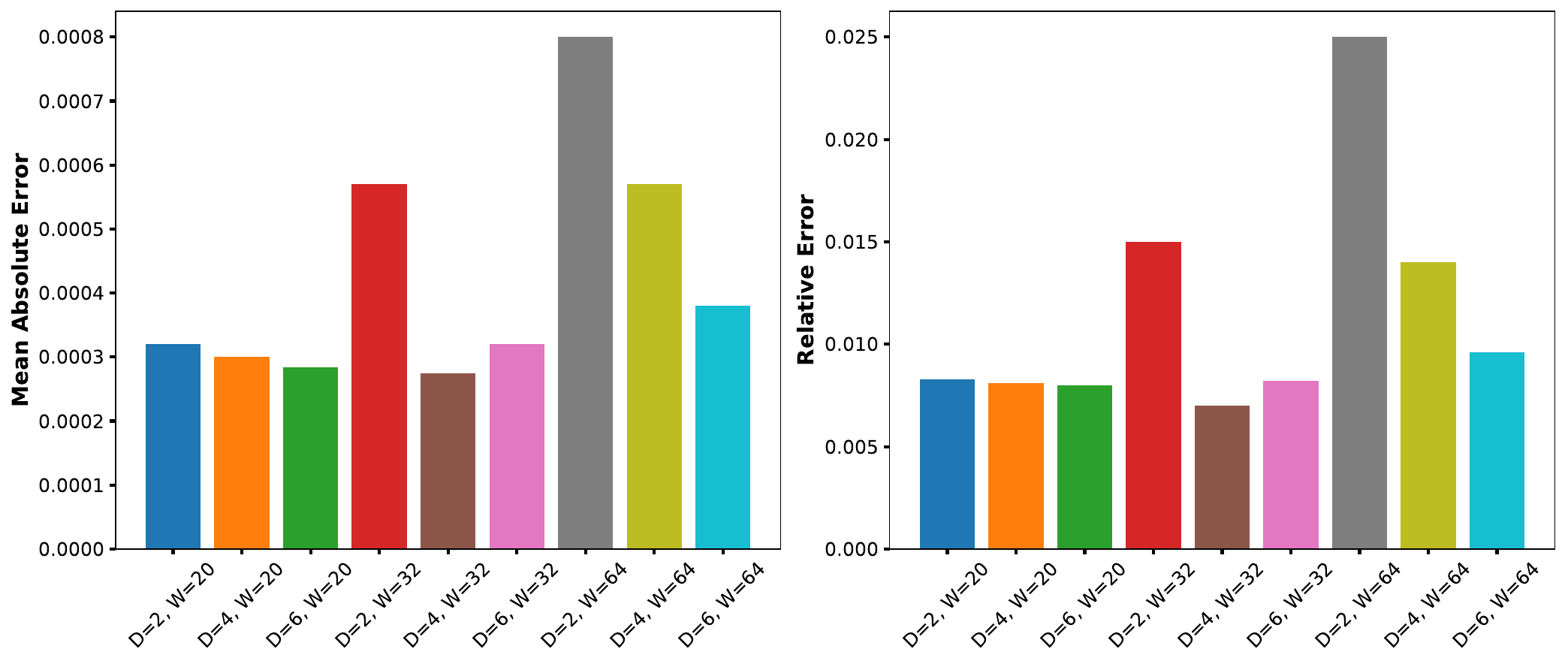}
 \caption{Mean Absolute Error (left) and Relative Error (right) for recovering $h(t)$ with 50 data points across nine architectures (depth $D$, width $W$). The network with $D=4,\,W=32$ yields the lowest errors }
\label{MAE_model3_1}
\end{center}
\end{figure}

\begin{figure}[H]
\begin{center}
\includegraphics[width=8in, height=3.0in, angle=0]{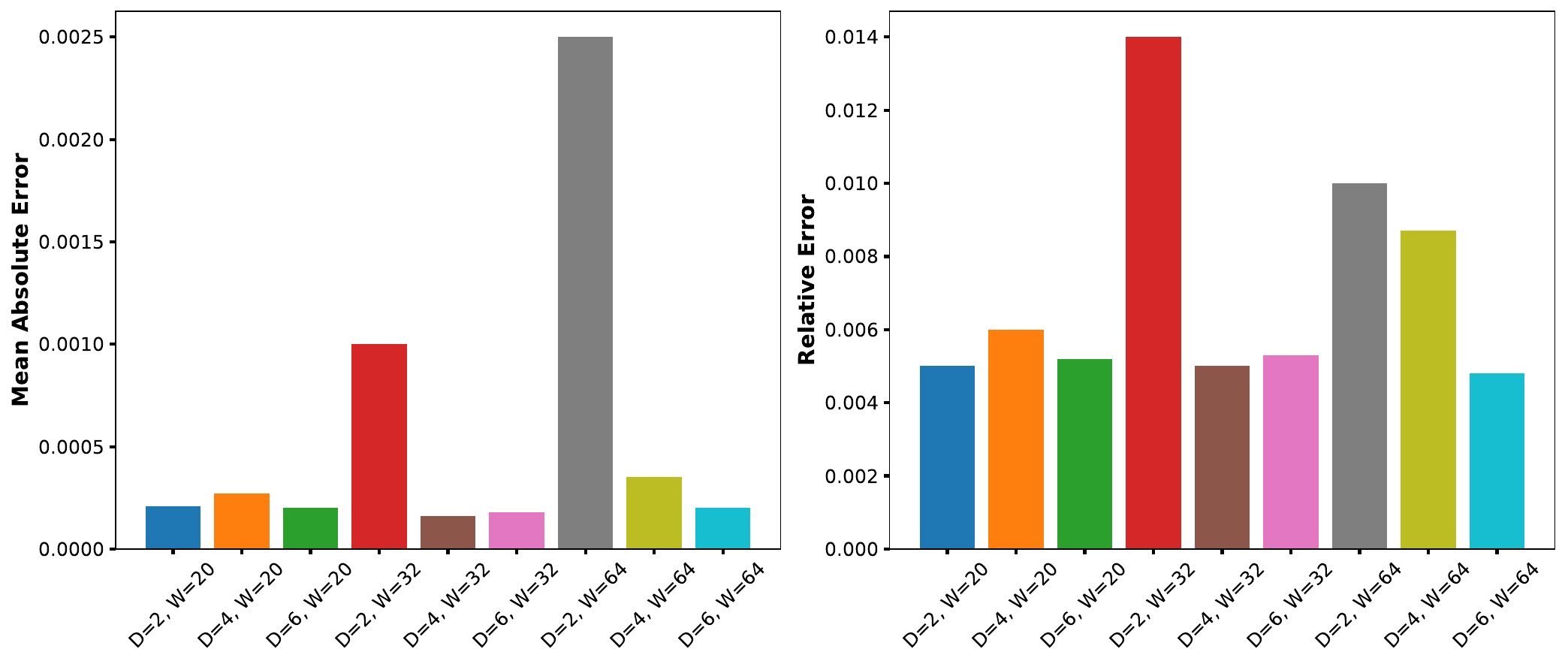}
 \caption{Mean Absolute Error (left) and Relative Error (right) for recovering $i(t)$ with 50 data points. Lowest MAE occurs at $D=4,\,W=32$ and lowest RE at $D=6,\,W=64$; shallow and wide models ($D=2,\,W=32/64$) show the largest errors. }
\label{MAE_model3_2}
\end{center}
\end{figure}

For missing physics identification, we have two unknown functions $h(t)$ and $i(t)$ to be discovered. The CBINN model predicted solutions and exact solutions is shown in Figure \ref{fig2_model3} using just 30 measurements for each state variables.  We clearly see that the unknown functions $h(t)$ and $i(t)$ are predicted exactly by the CBINN model. The MAE, RMSE, and RE for different data sizes is presented in Table \ref{table2_model3}. The performance of the model on noisy data is presented in Table \ref{table3_model3}.  We find that CBINN framework achieves great accuracy even with smaller datasets or noisy data. The performance of the model in discovering $h(t)$ and $i(t)$ with different depth and width of the network are illustrated in Figures \ref{MAE_model3_1}, \ref{MAE_model3_2}.

\begin{table}[h!]
\centering
\captionsetup{justification=centering}
\caption{Unknown term discovery for model (\ref{eq:T1})–(\ref{eq:A1}).\\
CBINNs performance via MAE, RMSE, RE, for different numbers of data points for both \(h(t)\) and \(i(t)\).}
\renewcommand{\arraystretch}{1.3}
\begin{tabular}{|c||c|c|c||c|c|c|}
\hline\hline
\multirow{2}{*}{\textbf{Data points}} 
  & \multicolumn{3}{c||}{\(\mathbf{h(t)}\)} 
  & \multicolumn{3}{c||}{\(\mathbf{i(t)}\)}  \\ 
\cline{2-7}
& \textbf{MAE} & \textbf{RMSE} & \textbf{RE ($\%$)} 
& \textbf{MAE} & \textbf{RMSE} & \textbf{RE ($\%$)}  \\ 
\hline\hline
200  
  & $2.3\times10^{-4}$ & $3.1\times10^{-4}$ & 0.55
  &    $1.0\times10^{-4}$       & $4.5\times10^{-4}$  & 0.4
   \\

\hline
100  
  & $2.76 \times10^{-4}$ & $3.7 \times10^{-4}$ & 0.8
  &  $1.6\times10^{-4}$ & $5.1\times10^{-4}$ &  0.5
   \\
\hline
50   
  & $5.5\times10^{-4}$ & $6.6\times10^{-4}$ & 1.4
  & $3.4 \times10^{-4}$ & $6.2\times10^{-4}$ & 1.1 
   \\
\hline
30   
  & $6.5\times10^{-4}$ & $8.4\times10^{-4}$ & 2.8
  & $5.3\times10^{-4}$ & $8.7\times10^{-4}$ & 2.5
  \\
  \hline
10   
  & $1.2\times10^{-3}$ & $3.2\times10^{-3}$ & 6.2
  & $3.2\times10^{-3}$ & $2.4\times10^{-3}$ & 7.1
   \\
\hline\hline
\end{tabular} \label{table2_model3}
\end{table}

\begin{table}[H]
\centering

\captionsetup{justification=centering}
\caption{Unknown term discovery for model (\ref{eq:T1})–(\ref{eq:A1}).\\ CBINNs performance via MAE, RMSE, and RE for different noise levels with 50 data points. }
\renewcommand{\arraystretch}{1.3}
\begin{tabular}{|c||c|c|c||c|c|c|}
\hline\hline
\multirow{2}{*}{\textbf{Noise Level}} 
  & \multicolumn{3}{c||}{\(\mathbf{h(t)}\)} 
  & \multicolumn{3}{c||}{\(\mathbf{i(t)}\)}  \\ 
\cline{2-7}
& \textbf{MAE} & \textbf{RMSE} & \textbf{RE ($\%$)} 
& \textbf{MAE} & \textbf{RMSE} & \textbf{RE ($\%$)} \\ 
\hline\hline
1\% 
  & $5.75\times10^{-4}$ & $8.5\times10^{-4}$ & 1.8
  &    $4.2\times10^{-4}$       & $7.6\times10^{-4}$  & 2.0
   \\

\hline
2\%  
  & $6.8 \times10^{-4}$ & $9.6 \times10^{-4}$ & 3.2
  &  $4.5\times10^{-4}$ & $8.2\times10^{-4}$ &  2.45
  \\
\hline
5 \%  
  & $1.45\times10^{-3}$ & $2.45\times10^{-3}$ & 8.3
  & $1.34 \times10^{-3}$ & $2.37\times10^{-3}$ & 7.2
  \\
\hline
10\%   
  & $2.1\times10^{-3}$ & $5.1\times10^{-3}$ & 12.5
  & $1.5\times10^{-3}$ & $3.2\times10^{-3}$ & 9.3
 \\
\hline\hline
\end{tabular} \label{table3_model3}
\end{table}

\section{Discussion and Conclusion}

Tumor-immune interactions are intrinsically multi-scale, nonlinear, and highly context dependent, so any compact mathematical model is necessarily a simplified description of a much richer biological reality.  Classical ODE models (for example, effector-tumor cells models discussed in \cite{kirschner1998modeling, kuznetsov1994nonlinear}) capture several qualitative behaviors observed experimentally but typically omit mechanisms such as spatial heterogeneity, evolving immune phenotypes, micro-environmental feedback, stochastic regulatory pathways, and other micro-environmental effects that can substantially alter dynamics in practice \cite{jarrett2018mathematical, wang2024agent, sardar2025stochastic}.  As a result, some terms in the true governing equations are often unknown or effectively absent from the working model; this observation motivates the gray-box (missing-physics) identification problem, namely the systematic discovery or correction of missing model terms from the data.  Hybrid approaches that combine mechanistic ODE structure with data-driven correction terms (physics-informed or structure-preserving machine learning) have recently been shown to be a promising route for recovering missing dynamics in biological systems, because they balance interpretability with flexibility and can recover corrective terms that are plausible \cite{yazdani2020systems, ahmadi2024ai, brunton2016discovering}.\\

Parameter estimation in tumour–immune models is particularly challenging because these models typically contain a large number of parameters, only a subset of model variables are measurable, and available experimental data are often sparse, irregularly sampled, and corrupted by noise. Classical methods such as least squares or standard Bayesian inference often require large datasets and careful initialization, and can be sensitive to noise and model stiffness. Physics-Informed Neural Networks (PINNs) address these issues because they (i) embed known mechanistic constraints directly into the loss function, reducing the effective search space for parameters, (ii) allow simultaneous estimation of unknown parameters and discovery of unmodeled terms from sparse/noisy observations using automatic differentiation, and (iii) regularize solutions toward physically consistent behaviors which improves robustness to noise and ill-posedness.  In \cite{raissi2019physics}, the authors introduce PINNs in the context of solving two main classes of problems: data-driven solution and data-driven discovery of partial differential equations. The effectiveness of the model is demonstrated through a collection of classical problems in fluids, quantum mechanics, and reaction–diffusion systems.  The authors in \cite{yazdani2020systems, ahmadi2024ai} developed a systems-biology-informed deep learning algorithm to infer the unknown parameters and missing terms.  Physics-informed neural networks are developed to simulate the incompressible Navier-Stokes equations in \cite{jin2021nsfnets}. Applications of PINNs to various prototype heat transfer problems is presented in \cite{cai2021physics}. \\

In this work, we present a cancer biology informed neural network (CBINN) framework to address the problems of parameter inference and missing physics identification in tumor-immune interaction models. Our approach successfully leverages partial system knowledge combined with limited, noisy experimental data to infer unknown parameters and the missing components within complex tumor-immune interaction models. This capability is crucial given the inherent biological complexity and experimental constraints in capturing all relevant mechanisms in the dynamics and complex tumor microenvironment. To evaluate the effectiveness and robustness of the CBINN framework, three compartmental models are selected. These models span a diverse set of dynamical behaviors, damped oscillations, sustained periodicity, and rapid initial growth leading to steady-state equilibrium which reflect common patterns observed in tumor-immune dynamics. The consistent performance of the proposed framework across these different regimes underscores its robustness and generalizability. Importantly, the developed model can discover the parameters and the missing physics accurately even when data are sparse and noisy, which is often the case in biological experiments. Moreover, the model is also able to accurately infer the dynamics of experimentally unobserved components.  This ability of the proposed model is important because it helps to better understand the complex systems where some processes or biological mechanisms are not fully known.\\

Overall, this study demonstrates the potential of neural network model to bridge the gap between mechanistic modeling and data-driven discovery in cancer modeling, offering a valuable tool for researchers to improve understanding of tumor-immune interactions and to guide experimental design and therapeutic strategy development. The proposed framework can be used for many physical systems to identify the mathematical expressions for the unknown parts when only partial knowledge of the physics is available. In this study, we have restricted our analysis to ODE based models. However, real tumor microenvironment often involve spatial heterogeneity and diffusion driven processes, which can be more accurately represented using partial differential equation (PDE)-based models. Extending our framework to PDE-based formulations is an important direction for future work, as it would allow for a more comprehensive representation of spatial-temporal dynamics. In this work, we used the Fisher Information Matrix (FIM) to approximate confidence intervals for the CBINN based parameter estimates. This approach relies on assumptions of estimator unbiasedness and normality and can be challenging in partially observed nonlinear models \cite{tangirala2018principles}. More advanced approaches such as Bayesian PINNs \cite{yang2021b}, likelihood profiling, or bootstrap-based methods \cite{joshi2006exploiting, foo2009multi} could be employed to obtain more accurate and robust uncertainty quantification. In \cite{ahmadi2024ai}, the authors have demonstrated the use of symbolic regression (SR) techniques, such as PySR and gplearn, in addition to the PINN architectures for addressing gray-box identification problems. Inspired by this approach, we also plan to incorporate SR methods in future studies to enhance model interpretability and improve the discovery of missing terms in the governing equations. \\

\noindent
\textbf{CRediT authorship contribution statement}\\
\noindent
BC developed the models, conducted the analysis, and wrote the article.
BVRK contributed in conceptual discussions, organization, planning, supervision, and corrections.\\

\noindent
 \textbf{Ethics Statement}  \\

This research does not require an Ethics statement. No direct experiments with human or animal subjects were conducted, and the
data used are publicly available or have been previously published
in peer-reviewed scientific literature. Therefore, no ethical committee
approval is needed, and the privacy rights of human subjects are not
compromised.\\

\noindent
 \textbf{Conflict of Interest:} Authors declare no conflict of interest.\\
 
\noindent
\textbf{Funding:} Not Applicable\\

\noindent
 \textbf{Data Availability:} Not Applicable \\

\noindent
\textbf{Use of AI tools Declaration}
The authors declare that no AI tools were used in the creation of this article.

\end{document}